\documentclass[sigconf]{acmart}
\AtBeginDocument{%
  }

\setcopyright{acmlicensed}
\copyrightyear{2026}
\acmYear{2026}
\acmDOI{XXXXXXX.XXXXXXX}
\acmConference[ICSE'26]{48th International Conference on Software Engineering}{April 12--18,
  2026}{Rio de Janeiro, Brazil}
\graphicspath{{./images/}}
\DeclareGraphicsExtensions{.pdf,.jpeg,.png,.eps}

\usepackage{caption}
\usepackage{subcaption}
\usepackage{enumerate}
\usepackage{booktabs}
\usepackage{algorithm}
\usepackage{algpseudocode}
\usepackage{hyperref}
\usepackage{nameref}
\usepackage{multirow}

\begin{document}

%%
%% The "title" command has an optional parameter,
%% allowing the author to define a "short title" to be used in page headers.
\title{SwarmUpdate: Hierarchical Software Updates and Deep Learning Model Patching for Heterogeneous UAV Swarms}

%%
%% The "author" command and its associated commands are used to define
%% the authors and their affiliations.
%% Of note is the shared affiliation of the first two authors, and the
%% "authornote" and "authornotemark" commands
%% used to denote shared contribution to the research.
% \author{Anonymouse}
% % \authornote{Both authors contributed equally to this research.}
% \email{}
% \orcid{1234-5678-9012}
\author{Lin Geng}
% \authornotemark[1]
\email{20lg5@queensu.ca}
\affiliation{%
  \institution{Queen's University}
  \city{Kingston}
  \state{Ontario}
  \country{Canada}
}
\author{Hao Li}
% \authornotemark[1]
\email{hao.li@queensu.ca}
\affiliation{%
  \institution{Queen's University}
  \city{Kingston}
  \state{Ontario}
  \country{Canada}
}
\author{Sidney Givigi}
% \authornotemark[1]
\email{sidney.givigi@queensu.ca}
\affiliation{%
  \institution{Queen's University}
  \city{Kingston}
  \state{Ontario}
  \country{Canada}
}
\author{Bram Adams}
% \authornotemark[1]
\email{bram.adams@queensu.ca}
\affiliation{%
  \institution{Queen's University}
  \city{Kingston}
  \state{Ontario}
  \country{Canada}
}

%%
%% By default, the full list of authors will be used in the page
%% headers. Often, this list is too long, and will overlap
%% other information printed in the page headers. This command allows
%% the author to define a more concise list
%% of authors' names for this purpose.
% \renewcommand{\shortauthors}{Anonymouse}

%%
%% The abstract is a short summary of the work to be presented in the
%% article.
\begin{abstract}
% Heterogeneous unmanned aerial vehicle (UAV) swarms consist of drones with varying roles, hardware, and software needs, presenting unique challenges in maintaining synchronized software updates. While software update strategies have been extensively studied for homogeneous UAV swarms, there is no prior research on updating software in heterogeneous UAV swarms. In this paper, we adapt and evaluate three update strategies originally designed for homogeneous swarms to the heterogeneous context: (1) an auction-based strategy, (2) a peer-to-peer strategy, and (3) a file replication strategy. We propose a benchmark for evaluating these strategies, measuring key factors such as update efficiency, scalability, and fault tolerance.

Heterogeneous unmanned aerial vehicle~(UAV) swarms consist of dozens to hundreds of drones with different roles and varying hardware and software requirements collaborating towards a shared mission. While traditional approaches for synchronized software updates assume swarms to be unstructured and homogeneous, the heterogeneous nature of modern swarms and the emerging need of drones to update their deep learning (perception) models with new objectives or data as a mission unfolds, has made efficient software update methods crucial for swarms to adapt to dynamic environments. 
To address these challenges, we introduce the SwarmUpdate framework for software updates in heterogeneous UAV swarms, composed of two key components: SwarmSync and SwarmModelPatch. SwarmSync is a hierarchical software update synchronization strategy to distribute a software update to the right subset of drones within a swarm, while SwarmModelPatch is a deep learning model patching method that reduces the size of a (deep learning model) update by only allowing some layers of the model to be updated (freezing the other layers). % while minimizing degradation in model accuracy.
In this paper, we systematically evaluate the performance of SwarmSync through large-scale simulations in the ARGoS swarm simulator, comparing SwarmSync to auction-based~(SOUL) and gossip-based rebroadcasting~(Gossip) baselines, and SwarmModelPatch to a non-incremental model patching strategy. %We analyze key performance metrics, including update convergence speed, transmission overhead, and resilience to communication failures across different swarm sizes.
Our results show that SwarmSync achieves up to 78.3\% faster update convergence compared to SOUL, and up to 47.7\% faster convergence compared to Gossip, while maintaining reasonable overhead, making it well-suited for large-scale swarm deployments. Furthermore, %we investigate the impact of adjusting the number of frozen layers in SwarmLLT in UAV swarms. O
our findings show that freezing seven out of eight layers of the model can reduce the update size by 73.3\%, speeding up the update process by an average of 72.2\% and reducing transmission overhead by an average of 74.3\%, at the expense of a drop of 5.1\% in overall accuracy (from 72.7\% down to 67.6\%). 

% \bram{are the numbers here in abstract median, average, ...?}

\end{abstract}

%%
%% The code below is generated by the tool at http://dl.acm.org/ccs.cfm.
%% Please copy and paste the code instead of the example below.
%%


%%
%% Keywords. The author(s) should pick words that accurately describe
%% the work being presented. Separate the keywords with commas.
%% A "teaser" image appears between the author and affiliation
%% information and the body of the document, and typically spans the
%% page.

% \received{20 February 2007}
% \received[revised]{12 March 2009}
% \received[accepted]{5 June 2009}

%%
%% This command processes the author and affiliation and title
%% information and builds the first part of the formatted document.

% Remember that if accepted, for the camera-ready version you will have to remove these commands.
\setcopyright{none} % to remove the copyright notice
\settopmatter{printacmref=false} % to remove the ACM Reference Format
\renewcommand\footnotetextcopyrightpermission[1]{}

\maketitle
    
\newcommand{\rqone}{What is the update synchronization efficiency of SwarmSync in a heterogeneous UAV swarm?}
\newcommand{\rqtwo}{What is the transmission overhead of SwarmSync in a heterogeneous UAV swarm?}
\newcommand{\rqthree}{What is the trade-off between efficiency, overhead and accuracy when updating a model with SwarmModelPatch?}

\section{Introduction}\label{sec:introduction}

Autonomous unmanned aerial vehicle~(UAV) swarms have emerged as a promising technology for accomplishing complex missions without relying on a central server~\cite{st2020design}. The last decade, research on swarms started focusing on heterogeneous UAV swarms comprising multiple UAV types, each designed with distinct roles and capabilities~\citep{pinciroli2016buzz}.
Compared to homogeneous swarms, heterogeneous UAV swarms offer enhanced scalability and performance, capable of addressing complex and diverse task requirements~\cite{varadharajan2024hierarchies, varadharajan2020swarm, ducatelle2010cooperative, kaminka2025heterogeneous}. 

An effective software update framework is crucial for swarms, as they often operate in complex field work where direct connectivity is limited or unavailable, such as search-and-rescue or remote surveillance missions, in mission-critical situations~\cite{xing2022reliability}. Recalling an entire swarm just to update their control program is not feasible in such settings. This is why prior research has proposed several swarm software update strategies~\cite{Varadharajan2018,al2022study}. For example, over-the-air programming (OTAP)~\citep{brown2013software, Varadharajan2018} allows updates to homogeneous swarms by rebroadcasting the update once it has been received by an individual drone. Software updates in swarms face challenges, such as bandwidth constraints, computational and storage constraints, and communication failures.% reducing the usability of simple adaptation from other software deployment scenarios.  \bram{mention resource-constrained devices and any other important challenges of swarm software updates}%but does not address the unique needs of heterogeneous swarms where UAVs have different update requirements. 

Unfortunately, existing software update strategies for swarms all assume (1) homogeneous UAV swarms (i.e., all UAVs have identical roles and update requirements), and (2) software updates comprising typical code and/or data payloads instead of deep learning~(DL) model updates. Heterogeneous swarms introduce complications such as multiple software versions and configurations, different hardware and storage capabilities, and incompatibility with certain software. Furthermore, UAVs increasingly adopt DL models, such as convolutional neural networks (CNNs), for perception tasks, which require updates during long-running missions to prevent concept drift~\cite{sato2019continuous, hashmani2019accuracy}. Without reliable and efficient updates, and effective update strategies able to deal with UAVs' limited resources, UAVs risk mission failure from outdated software or DL models.%existing approaches rarely consider scenarios involving DL models, despite their increasing adoption and the complexities in updating the model within resource-constrained UAV systems.

To address these challenges, we introduce a framework tailored to heterogeneous UAV swarms called \textbf{SwarmUpdate}, which has two components: \textbf{SwarmSync} and \textbf{SwarmModelPatch}. SwarmSync is a hierarchical software update synchronization strategy designed for heterogeneous UAV swarms that categorizes UAVs into sub-swarms to organize the distribution of software patches.
%Hierarchical systems have better version control than egalitarian systems as they only distribute the update to those in need.
% that balances structured leader-based updates with distribution strategies, specifically built for larger update sizes such as DL model updates. This approach is benchmarked against auction-based~(SOUL) and gossip-based~(Gossip) to evaluate the efficiency, reliability, and overhead. 
SwarmModelPatch is a DL model patching method that focuses on reducing the size of a model update while minimizing degradation in model accuracy by selectively freezing model layers before tuning. This paper also systematically evaluates SwarmUpdate through comprehensive simulation experiments, comparing its performance against existing baseline methods.

To illustrate the potential impact and importance of our study, consider the following real-world scenario. A surveillance UAV swarm is tasked with patrolling remote mountainous areas far beyond the range of infrastructure support. When this swarm encounters sudden environmental changes~(e.g., unexpected snowfall), the DL models on the UAVs' perception systems may become unreliable, severely affecting mission outcomes. The mission cannot afford downtime for updates, yet adaptability is crucial. Instead, the swarm could send a small number of UAVs back carrying any relevant new data, notifying headquarters that an update is required to maintain functionality, or communicate new data back through peer-to-peer communication links between UAVs all the way back to base. In response, headquarters trains the existing model with the new data, then sends an ``Updater'' UAV back to the swarm, responsible for distributing the updated model to any UAV that requires an update using the SwarmSync synchronization approach. Once a UAV receives the update, it leverages the SwarmModelPatch model patching method to incrementally update its DL model.

The contributions of this paper are summarized as follows:
\vspace{-\topsep}
\begin{itemize}
\item SwarmSync: A software update synchronization strategy for heterogeneous UAV swarms, leveraging the hierarchical structure of swarms to distribute software patches.
\item SwarmModelPatch: The first DL model patching method designed for UAV swarms to train DL models to adapt to new data.%, create and apply patches, and reduce update size with minimal accuracy loss by adjusting the number of frozen model layers.
% \item \bram{is this a real contribution, or basic argos?} A simulation and systematic evaluation for software update frameworks in heterogeneous UAV swarms.
\item A replication package contains simulation environment setups, scripts, data, and DL models at \url{https://anonymous.4open.science/r/SwarmUpdate-860D}.%\bram{we should mention the replication package comprising scripts and data, putting it on a temporary URL (dropbox?) for now, making sure it's double-blind (i.e., anonymous)}
% \item A Deep Learning update framework SwarmEvolve focusing on the efficiency of peer-to-peer updates in a heterogeneous UAV swarm.
% \item A systematic evaluation in a simulation environment to compare SwarmSync with auction-based and gossip-based strategies, showing SwarmSync accomplishes 80\% faster convergence against auction-based technique and 56\% faster convergence against gossip-based techniques.
% \item We adapt SOUL and Gossip to a heterogeneous setting, benchmarked their efficiency against our strategy.
% \item We propose and benchmarked SwarmSync against SOUL and gossip to ensure it has good performance.
% \item A systematic evaluation for the effectiveness of SwarmLLT and the effect of patch sizes on the update strategies.
\end{itemize}
\section{Related Work}\label{sec:related_work}

\textbf{Software Update Strategies.}
Existing research on software update strategies primarily target homogeneous swarms~\citep{Varadharajan2018, al2022study}. For instance, \citet{Varadharajan2018} propose an update protocol incorporating rebroadcasting and retransmission through requests, while \citet{al2022study} focus on security threats when updating a swarm. However, these studies do not address the complexities inherent in updating heterogeneous swarms, where different UAVs have distinct roles and diverse update requirements. Additionally, traditional software patching usually replaces the older version, which is unsuitable for resource-constrained UAV swarms with DL models. SwarmUpdate only transmits the differences between model versions. Moreover, by freezing earlier layers, SwarmUpdate minimizes update size and bandwidth usage.
% To the best of our knowledge, our paper proposes the first software update strategies for heterogeneous swarms, leveraging a hierarchical structure to enhance efficiency.

% Traditional software patching usually replaces the older version, however, in bandwidth-limited environments, transmitting and replacing the entire file causes redundant and wasteful transmissions. Techniques such as bsdiff, where only the difference of the two versions are recorded, are designed to only transmit the changed portion of the software, lowering the patch size and reducing the overall bandwidth requirements. Utilizing this idea, we only transmit the difference between the two models to reduce patch size, and we attempt to compact the differences into just a few layers by freezing earlier layers.
% \bram{we should also make clear that the traditional "patch" technique is "replace", while we claim that for models an "incremental replace" based on freezing layers is better; the "replace" patching approach is our baseline}

Since egalitarian, homogeneous swarms can be considered as a Peer-to-Peer~(P2P) system, swarm researchers have explored various software update strategies for P2P systems~\cite{busnel2007gcp}. For example, a widely used protocol for P2P system software updates is Gossip-based protocols, \citet{busnel2007gcp} use gossip-based protocols for distributing software updates for wireless sensor networks. However, gossip-based protocols often suffer from redundant messaging, inconsistencies, and lower performance in sparse connections~\cite{pinciroli2016tuple}.
\noindent\textbf{Deep Learning Model Updating.}
Updating deep learning (DL) models is a critical challenge for updates in UAV swarms, as DL models are usually large and bandwidth-consuming. Researchers have proposed various techniques to mitigate the need for full retraining, such as transfer learning~\cite{weiss2016survey} and lifelong learning~\cite{olewicki2024costs,parisi2019continual}. The idea of lifelong learning can be traced back to~\citet{mccloskey1989catastrophic} in 1989, which proposes training a model incrementally on new data. Recently, \citet{olewicki2024costs} explore cost-efficient lifelong learning techniques to reduce computational overhead compared to retraining. To avoid catastrophic forgetting, a replay buffer is used containing a portion of old training data. Transfer learning leverages pre-trained models for adaptation to new but related tasks~\cite{weiss2016survey}, lowering the computational resources needed to train. However, these studies primarily focus on improving accuracy or minimizing training resources~\cite{romiti2022realpatch,goel2020model,ilharco2022patching}, rather than reducing the size of model updates. To the best of our knowledge, our paper is the first to propose a DL model patching method that aims to minimize model changes and hence to reduce model patch sizes.

While compression techniques like pruning~\cite{he2023structured} and quantization~\cite{goel2020model, zhou2018adaptive, rokh2023comprehensive} reduce overall model size and computational cost, they do not specifically focus on minimizing differences between model versions to optimize patch sizes. Pruning removes less important neurons post-training, whereas quantization reduces memory and computation by lowering weight precision (e.g., from 32-bit float to 8-bit int). However, neither explicitly targets reducing the model update size between versions.

% While compression\cite{choudhary2020comprehensive} techniques such as pruning\cite{he2023structured}, and quantization\cite{goel2020model,zhou2018adaptive,rokh2023comprehensive} exists to lower the size of a model, they do not target the difference between two versions of a model to reduce the size of patches. Model Pruning focuses on reducing the size and computational costs of deep learning models by removing less important neurons, typically used after a model is trained. Model quantization uses reduces memory demand and computational demand by converting weights into lower precision data types (e.g. from 32-bit float to 8-bit int), while this technique could be applied to reduce the patch size, this technique does not target reducing the difference between two versions of a model. \bram{what about techniques like quantization, distillation, etc.? need to mention those here somewhere}

% \vspace{0.5em}
\noindent\textbf{Software Engineering for UAV Systems.}
Software engineering researchers have studied different aspects of UAV systems such as dynamic updates, robustness, and vulnerabilities. For example, \citet{nahabedian_dynamic_2020} introduce an automated dynamic controller synthesis approach to facilitate efficient updates of UAV controllers, validated in UAV surveillance systems. To enhance the robustness of UAV systems, \citet{wang_uavanomaly_2024} present a method for detecting anomalies from UAV logs, while \citet{han_uavconfig_2022} implement a learning-guided search framework to detect vulnerabilities in UAV configuration modules. In addition, \citet{jung_uavbugs_2021} propose a dedicated debugging system for identifying and resolving configuration-related bugs in UAV swarms. 

Regarding system interpretability and safety, \citet{ataiefard_deep_2022} propose a hybrid deep neural network method to infer internal states of UAV autopilots by analyzing input-output signal data. To ensure safety in UAV software, \citet{liang_uavsafety_2021} empirically evaluate the usage of bounding functions within open-source UAV software frameworks to understand safety-critical concerns. Furthermore, \citet{vierhauser_interlock_2021} introduce a framework for interlocking Safety Assurance Cases (SACs) to improve operational accountability, where each UAV provides a pluggable SAC demonstrating compliance with infrastructure-defined safety constraints.

None of the above work targets the unique needs and complications a heterogeneous UAV swarm faces to update a DL model in-field, nor attempt to improve the efficiency of software updates in UAV swarms.

\section{SwarmUpdate}\label{sec:methodology}

This section presents the problem definition of this study and the two main components of SwarmUpdate
%for updating software in heterogeneous UAV swarms
: a hierarchical software update synchronization strategy called SwarmSync and a DL model patching technique called SwarmModelPatch. %We propose a hierarchical update strategy designed to accommodate heterogeneous UAV swarms while ensuring fast convergence with low communication overhead.

% \subsection{Update Strategies}

% We implement and compare three update strategies

\subsection{Problem Definition}
% Traditional strategies, such as centralized broadcasting, or gossip-based rebroadcasting, is inefficient in large-scale deployments due to network congestion, high transmission overhead, and susceptibility to packet loss. UAVs also often have limited computational resources and intermittent connectivity, making efficient and scalable software updates a challenging task.

Consider a UAV swarm as a set \( S = \{ u_1^{t_1}, u_2^{t_2}, \ldots, u_n^{t_n} \} \), where each UAV \( u_i^{t_i} \) belongs to a type \( t_i \), runs software with version \( v_i^c \), and has a given communication range \( r_i \). A software version \( v^c \) is compatible with UAVs whose type \( t \) satisfies \( t \in c \), where $c$ are the compatible types for that software update, $c=\{t_j | t_j\ is\ compatible\ with\ v^c\}$. UAVs incompatible with the update (i.e., \( t \notin c \)) are excluded from receiving it. Let \( S^c \subseteq S \) represent the subset of UAVs compatible with update \( v^c \). The goal is to distribute the latest software version \( v_{\text{new}}^c \) to all compatible UAVs in \( S^c \), ensuring update coverage with minimal latency and overhead.

A subset \( S^m \subseteq S \) of UAVs are equipped with an imaging sensor whose input is fed to a DL CNN model $\mathcal{M}$ to perform perception tasks. Consider such a model $\mathcal{M}$ capable of classifying images into classes~(labels) $L=\{l_1,l_2 \ldots l_n\}$ with accuracy $Acc$. When a new class $l_{n+1}$ is introduced, the existing model must be updated to a new model version $\mathcal{M}^*$ capable of classification into $L^*=\{l_1,l_2 \ldots l_n,l_{n+1}\}$ with accuracy $Acc^*$. Updating the existing model from $\mathcal{M}$ to $\mathcal{M}^*$ generates a model patch $\mathcal{P}$ with size $s$. The goal is to minimize $s$ with minimal degradation in $Acc^*$.

% Let a UAV swarm be defined as a set of UAVs $S=\{u_1^t,u_2^t,...,u_n^t\}$ where each UAV $u_i^t$ is of type $t$. Each $u_i^t$ has a communication range $r_i$, and a software version $v_i^c$, where $c$ is the compatible UAV types for that software version. A UAV $u_i^t$ can get a software of version $v_i^c$, if $t\in c$. If $t \notin c$, then the UAV is not compatible with the update and will not receive the update. The subset of $S$ where $u_i^t | t \in c$ is represented as $S^c, S^c\subseteq S$. The object of this work is to ensure that all UAVs in $S$ receive the latest software update $v_{new}^c$, where $t\in c$, with minimal latency and overhead.

\begin{algorithm}[t]
\caption{SwarmSync's Updater}
\label{alg:Updater}
\begin{algorithmic}[1]
    \State \textbf{Signal:} Update is available
    % \State $num\_leaders \gets$ count leaders in swarm
    \State Wait until all leaders get in position \Comment{Until Time-out}% \bram{time-out possible here as well: might cancel overall update?}
    \For{each packet in packets}
        \Repeat
            \State Transmit current packet to all leaders
        \Until{ACK received from all leaders} \Comment{Until Time-out}% \bram{what happens? the whole update is aborted? only that leader's update is aborted? ...}
    \EndFor
    \If{Time-out}
        \State \textbf{Signal} Reappoint Leader
        \State \textbf{Restart} Update Process
    \EndIf
    \State Wait until convergence signals received from all leaders
    % \If{signal\_converged from all leaders}
    %     \State Converged
    % \Else
    %     \State Wait for convergence
    % \EndIf
\end{algorithmic}
\vspace*{-0.2em}
\end{algorithm}

\begin{algorithm}[t]
\caption{SwarmSync's Leader}
\label{alg:Leader}
\begin{algorithmic}[1]
    \State \textbf{Receive} signal from Updater
    \State $PatchSize \gets$ from signal, propagate signal to swarm
    \State Move toward and surround Updater% \bram{why needed, since already received initial signal? to optimize reception during the alg.2 process?}
    \State \textbf{Signal:} leader in position
    \State \textbf{Receive} packet from update
    \While{\#packets $<$ $Patch Size$}
        \State Wait to receive the packet \Comment{Until Time-out}% \bram{here as well, explain in text what happens during time-out}
        \If{packet received successfully}
            \State \textbf{Send} ACK
        \EndIf
        \If{Time-Out}
            \State \textbf{Abort} Update process, signal others to also abort
        \EndIf
        % \If{update packet not received}
        %     \State Add packet to storage
        %     \Repeat
        %         \State \textbf{Send} ACK
        %     \Until{new packet received}
        % \EndIf
    \EndWhile
    \State Return to respective sub-swarm
    % \State Broadcast update to followers
    \For{each packet in packets}
        \Repeat
            \State Transmit current packet to all followers
        \Until{ACK received from all followers}  \Comment{Until Time-out}
        \If{Time-out}
            \State \textbf{Remove} Unresponsive follower from follower list
            \State \textbf{Continue} Next packet
        \EndIf
    \EndFor
    \State \textbf{Signal:} Notify Updater that sub-swarm has converged
\end{algorithmic}
\vspace*{-0.3em}
\end{algorithm}

\begin{algorithm}[t]
\caption{SwarmSync's Follower}
\label{alg:Follower}
\begin{algorithmic}[1]
    \State \textbf{Receive} signal from Updater
    \State $PatchSize \gets$ from signal, propagate signal to neighbours
    \State Remain idle until leader returns with update% \bram{so the followers continue their mission, they do not move out of position for the update: make this clear in the text}
    % \State Wait for leaders to update
    % \State \textbf{Signal:} leader in position
    % \State \textbf{Receive} packet from update
    \While{\#packets $<$ $PatchSize$}
        \State Wait to receive the packet \Comment{Until Time-out}
        \If{Time-out}
            \State \textbf{Signal} Time-out, reappoint leader in the swarm
        \EndIf
        \If{packet received successfully}
            \State \textbf{Send} ACK
        \EndIf
        % \If{update packet not received}
        %     \State Add packet to storage
        %     \Repeat
        %         \State Send ACK
        %     \Until{new packet received}
        % \EndIf
    \EndWhile
    \State \textbf{Apply} Patch
    \State \textbf{Signal:} Notify leader completion of update% \bram{entire update or only synchronization part?}
\end{algorithmic}
\vspace*{-0.2em}
\end{algorithm}

\subsection{SwarmSync}

We propose SwarmSync to ensure efficient and reliable software update synchronization for heterogeneous UAV swarms, basically distributing a software update payload (whether it represents an update to a new model version $\mathcal{M}^*$, or more traditional code or data updates). SwarmSync is inspired by the scalable SWARM data replication method~\citep{Shen2015, mohammadi2019data}, where a designated supernode distributes data within server networks to other nodes with the same interest. In SwarmSync, the swarm is divided into smaller sub-swarms based on UAV types, where each sub-swarm consists of only one type of UAVs $t_i$ and appoints a leader~(either randomly or based on predefined rules, such as the lowest UAV identifier ID). Note that there might be multiple sub-swarms for a given UAV type $t_i$, to avoid too unbalanced sub-swarms. Updates propagate via a structured and hierarchical mechanism based on three distinct roles: Updater, leader, and follower. The ``Updater'' UAV is the one physically triggering the update process in close collaboration with the sub-swarms' ``leader'' UAVs. The latter then distribute updates to the other UAVs in their sub-swarm, i.e., the ``follower'' UAVs.

To ensure the guaranteed convergence of the swarm and prevent false positives, SwarmSync uses the Transmission Control Protocol~(TCP)~\cite{cerf1974protocol}. In TCP, every network packet sent requires an acknowledgement~(ACK) signal back before the sender moves on to the next packet. In the absence of an ACK signal, the sender keeps on re-sending the current packet. %To adapt to the simulation and a group setting, the sender will continue sending the same packet each control step until an ACK signal is received from all follower drones.
% resulting in an extra transmission and a one control step delay, the delay is a simulation specific limitation that is not present in real-world applications. 
Since each leader tracks its assigned followers, this method guarantees that all drones will receive all packets of the update, which is important as UAVs are often used in mission-critical or safety-critical situations~\cite{xing2022reliability}. That said, TCP's reliance on ACK signals introduces a potential deadlock risk if a UAV fails to respond. To mitigate potential deadlocks, SwarmSync implements a time-out mechanism: if a drone is unresponsive within a given time frame,%~(e.g., 20 control steps)%\bram{control steps belong to the empirical study, not the general algorithm section}
it is removed from the leader's follower list; if a leader is unresponsive, the sub-swarm will appoint a new leader in its place. %Given the importance of reliable convergence, some efficiency trade-offs generated from retransmission in TCP are acceptable compared to unstructured protocols like UDP. \bram{does not belong here, but in empirical study, although it's risky to mention since people might say the reliability property has not been evaluated}: 
Although a time-out did not occur during the simulation, preventative measures are essential for critical systems. 
% An example structure for SwarmSync is shown in \autoref{fig:SwarmSync_structure}.

% \begin{figure}[t]
%     \centering
%     \includegraphics[width=\linewidth]{images/SwarmSync_structure.pdf} 
%     \caption{An example structure formation for SwarmSync.}
%     \label{fig:SwarmSync_structure}
% \end{figure}
% The update process for each type is as follows:

\textbf{Updater.}
As presented in~\hyperref[alg:Updater]{Algorithm~\ref{alg:Updater}}, the Updater signals the swarm that an update is available. This signal contains metadata including version details and update packet size. The Updater then waits for the leaders' signal that indicates they are ready in position. The Updater sequentially sends update packets and waits for ACKs from all leaders before proceeding. After completing transmissions, the Updater remains idle and waits for convergence signals from the leaders. Once the Updater receives convergence signals from all leaders, the swarm is declared to have converged. As the Updater depends on the leaders in two scenarios: (1) waiting for all leaders to get in position before sending an update, and (2) waiting for ACK from all leaders before sending the next packet, a deadlock could happen if a leader becomes unreachable, and thus time-out is introduced for both scenarios. %To prevent deadlocks, a time-out mechanism is implemented in both scenarios. 
When a time-out happens, a signal is sent out to the swarm, notifying that the leader has been removed from the swarm. The sub-swarm of the missing leader then appoints a new leader and restarts the update process.% \bram{explain the time-out remediation strategies here, very important} 
% the first packet, then wait for ACK from all leaders, if an ACK is missing, the Updater will assume a packet loss has occurred, and retransmit the packet. When the Updater receives ACK from all leaders, the Updater will then move on to the next packet. This process is repeated until all packets for the update are sent. The Updater will then go idle and wait for convergence signal from the leaders. Once the Updater receives convergence signal from all leaders, the swarm is determined to be converged. 
% The pseudo-code for the Updater is presented in~\hyperref[alg:Updater]{Algorithm~\ref{alg:Updater}}.

\textbf{Leader.}
As shown in \hyperref[alg:Leader]{Algorithm~\ref{alg:Leader}}, when an update signal is received, either from the Updater or from an UAV propagating the update signal, the leader of each sub-swarm moves to the Updater's position and forms a predefined spatial formation~(shown in \autoref{fig:Example_formation}) to get in range to receive updates, temporarily leaving its default position.

Once in position, leaders signal their readiness to receive updates. They sequentially receive packets from the Updater and send ACK for each packet. After receiving all packets, leaders return to their respective sub-swarms, distributing the update packets to their followers using the same reliable transfer~(TCP) approach. Leaders notify the Updater when all their followers have completed their update. %The leader waits until all followers complete the update before applying the patch to itself, in prevention of cases where the update must be aborted. 
If a leader has no followers, the leader will immediately notify the Updater that the sub-swarm has converged. 

To prevent deadlocks, the leader also implements a time-out mechanism, in two scenarios. First, if the leader times out while receiving an update from the Updater, the leader assumes the Updater has a failure and the update cannot be completed. Since only the Updater possesses the update, the process is aborted, and an abort signal is sent by the leader and propagated to other UAVs in the swarm. Second, if the leader times out while waiting for ACKs from the followers, any non-responding follower is removed from the follower list and the update continues for remaining followers. %\bram{do the followers apply the patching approach before notifying the leader/Updater?} \bram{also explain what happens upon time-out}

% update the sub-swarm, following a similar strategy to the Updater. Once all followers has given ACK to all packets, the leader will signal that this sub-swarm has converged to the Updater.  The pseudo-code for the leader is presented in \hyperref[alg:Leader]{Algorithm~\ref{alg:Leader}}.

\textbf{Follower.}
As the drones have only a limited communication range $r_i$, peer-to-peer propagation of data in a swarm is essential in order to reach all UAVs. Shown in \hyperref[alg:Follower]{Algorithm~\ref{alg:Follower}}, UAV followers propagate the update signal obtained from their sub-swarm leader to all their neighbors to expedite the signal. While the leaders are getting the update from the Updater, the followers continue their usual activities but remain within the communication range of their leader's last location~(before moving to the Updater)
% in place, continuing any activity that does not require them to move and stopping any activity that requires movement\bram{they don't continue their usual activity?}
until the leader returns to the sub-swarm. The followers will send an ACK for each received packet to their leader. Once all packets are received, the followers will apply the patch and signal their leader about completion. Since the follower depends on the leader to both return and send update packets, a time-out mechanism is in place: the follower assumes that the leader is unreachable and sends a signal to reappoint a new leader for the sub-swarm.

%In ideal scenarios, if \( |S^c| \) is the number of UAVs requiring updates, and \( N \) is the maximum number of UAVs updated simultaneously per cycle \bram{what is a cycle? we're not yet talking about the empirical evaluation/simulation} from one Updater, the swarm update completion occurs after \( \left\lceil \frac{\ln(|S^c|)}{\ln(N)} \right\rceil \) cycles, as each leader can propagate the update to $N$ drones \bram{I don't understand the use of ln here}. 
In ideal scenarios, if \( |S^c| \) is the number of UAVs requiring updates, and \( N \) is the maximum number of UAVs updated simultaneously per cycle, where a cycle is the amount of time a drone require to update.
For swarm sizes $|S^c| \leq N^2$, SwarmSync can complete the update process in two update cycles. For swarms of size $|S^c|\leq N$, the Updater will be appointed as the temporary leader and all followers be updated directly by the Updater instead, completing the update process in one update cycle. For $|S^c|>N^2$, subsub-swarms can be created recursively to maintain the hierarchical structure of the swarm, where each follower of the sub-swarm can lead their own subsub-swarm and propagate the update. In practice, delays arising from positional adjustments or network conditions may increase this reported optimal complexity.
% SwarmSync can complete the update process in 2 update cycles for swarm sizes $u<=N^2$, where:
% $u$ represents the number of drones that require an update. $N$ is the maximum number of drones that a single Updater can reach in one cycle. In the case where $u>N^2$, subsub-swarms will be created to maintain the hierarchical structure of the swarm. As each leader can update $N$ drones, each update cycle $c$ can grow the total droned updated exponentially, meaning the number of drones can be updated for a given update cycle is measured by $N^c=u$. The required update cycles for the swarm to converge, in optimal settings, can be measured with $c = \left\lceil \frac{\ln(u)}{\ln(N)} \right\rceil$, real-world implementations can introduce delays that does not exist ARGoS.

\begin{figure}[t]
    \centering
    \includegraphics[width=\linewidth]{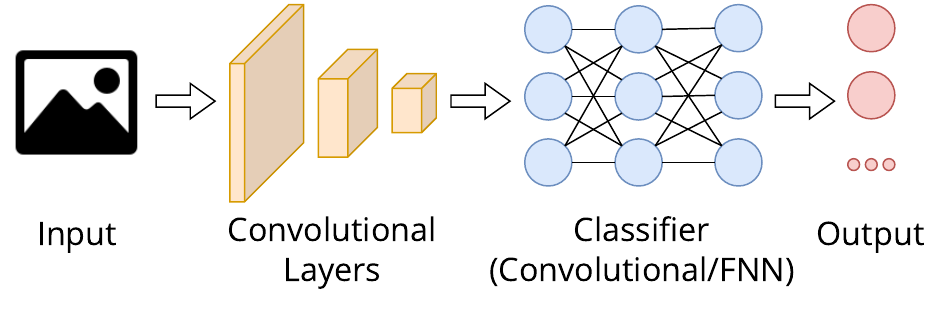} 
    \caption{Structure of a typical CNN model.}
    \label{fig:CNN}
\vspace*{-0.5em}
\end{figure}

\subsection{SwarmModelPatch}\label{sec:SwarmModelPatch}

\begin{algorithm}[t]
\caption{SwarmModelPatch, referring to \autoref{fig:CNN}}
\label{alg:Model}
\begin{algorithmic}[1]
    \State \textbf{Input:} Old CNN model $\mathcal{M}$, Replay Buffer Size $\beta$, Trainable Layers $x$ %\bram{is the approach unique to CNN, or it applies to any DL, we just only evaluated CNN?}
    \State \textbf{Output:} Updated model $\mathcal{M}^*$, Patch file  $\mathcal{P}$
    \vspace{5pt}
    
    \State \textbf{Prepare} training data:
    \State \quad $\mathcal{D} \gets \text{ReplayBuffer (old data)} \cup \text{New Data}$ %\bram{where is replay buffer updated for next time?}
    
    \State \textbf{Load} model $\mathcal{M}$
    \State \textbf{Reinitialize} classifier layer $\mathcal{C}$ in $\mathcal{M}$ %\bram{what does this mean? difference with M? it's classifier layer in M?}
    
    \State \textbf{Freeze} all layers except:  
    \State \quad - Last $x$ convolutional layer(s) \Comment{Adjust accordingly}
    \State \quad - Classifier $\mathcal{C}$
    
    \State $\mathcal{M}^*  \gets$ Train $\mathcal{M}$ on $\mathcal{D}$ \Comment{Update trainable parameters}
    % \State \quad \quad Update trainable parameters

    % \State \quad $\mathcal{M}^* \gets $ Updated model
    \State \textbf{Initialize} empty dictionary $\mathcal{P}$
    
    \State \textbf{Iterate through parameters of $\mathcal{M}^*$:}
    \For{\textbf{each} parameter $p$ \textbf{in} $\mathcal{M}^*$}
        \If{$p$ \textbf{not in} $\mathcal{M}$ }%\textbf{or} shape differs \textbf{or} values differ
            \State $\mathcal{P}[p] \gets \mathcal{M}^*[p]$
        \EndIf
    \EndFor
    
    \State \textbf{Save} $\mathcal{P}$ to patch file
    \State \textbf{Return} Updated model $\mathcal{M}^*$, Patch file $\mathcal{P}$
\end{algorithmic}
\vspace*{-0.2em}
\end{algorithm}

Once the payload of a software update is synchronized across all UAVs in a swarm, it needs to be installed. For traditional code/data updates, UAVs would typically ``replace'' the previous version's code/data. However, in the case of a deep learning model update, a naive ``replace'' strategy would require an entire new model version to be re-trained/fine-tuned, then sent in the payload of an update. This would not scale for typical swarm settings, where only intermittent peer-to-peer communication is available. %we propose SwarmModelPatch, which leverages lifelong learning techniques % Swarm Lifelong Learning Transfer~(), a method designed 
%to efficiently update a DL CNN model deployed on a UAV. UAV swarm operations frequently encounter dynamic environments which require periodic CNN model updates. 

To effectively deal with model updates and reduce the costs of retraining from scratch, we propose SwarmModelPatch, which integrates ideas from lifelong learning~\cite{olewicki2024costs,parisi2019continual}, transfer learning~\cite{cs231n_transfer_learning,neptune_transfer_learning_guide}, and incremental learning \cite{polikar2001learn++}. As shown in \hyperref[alg:Model]{Algorithm~\ref{alg:Model}}, SwarmModelPatch maintains a replay buffer holding representative samples from previously seen training data. When an environmental change occurs, the model update incorporates both the replay buffer and new data to prevent catastrophic forgetting~\cite{olewicki2024costs} and enable rapid adaptation.

% As many UAV rely on machine learning models to accomplish their task, it is important to study how to optimize the update process for ML models with challenges from UAV swarms.

% To reduce the costs of retraining from scratch, we integrate concepts from Lifelong Learning~\cite{olewicki2024costs,parisi2019continual}, Transfer Learning, and Incremental Learning. Specifically, we use a replay buffer to store parts of the old training data, we then train the model with a combination of old data and new data. 
% \bram{here we need to refer back to transfer learning and the layer-based architecture of DL models such as CNN to motivate the use of freezing} 
A typical CNN model is shown in \autoref{fig:CNN}, a CNN model usually consists of some trainable convolutional layers, untrainable pooling layers, and a classifier which can be either a feed-forward neural network~(FNN) or convolutional layer(s). The output of the classifier is used to determine the classification results. % \bram{this paragraph goes very quickly over things, jumps too fast}
To efficiently enable a given model to classify data according to newly introduced classes, SwarmModelPatch reinitializes the classifier layer~\cite{shen2021partial,fregier2021mind2mind}, while partially freezing the remaining layers. Once a new model version is trained, SwarmModelPatch generates a model patch representing the differences between the new and old model versions, essentially only containing the changed layers and metadata about these layers (and those they attach to), while ignoring the frozen layers. 

\hyperref[alg:Model]{Algorithm~\ref{alg:apply_patch}} shows the algorithm of SwarmModelPatch for applying a patch. If a parameter $p$ in patch $\mathcal{P}$ has the same shape as the corresponding parameter in model $\mathcal{M}$, it is added to $\mathcal{M}[p]$. However, if $p$ does not exist in model $\mathcal{M}$ or has a different shape, $\mathcal{M}[p]$ is replaced with $\mathcal{P}[p]$.  
% \bram{\autoref{alg:Model} only mentions patch generation, not patch application}

% To reduce the update size, we freeze all layers except the last convolution layer and perform transfer learning to adapt the model to the new update.

% We refer to this process LifelongIncremental. \hyperref[alg:Model]{Algorithm~\ref{alg:Model}} presents the pseudo code for the model update.

% After a new model has been obtained, we create a patch with both versions of the model, this patch should account for any new classes being added. We acknowledge that strategies such as quantization could be used to compress the patch even further, but we decided to keep the patch unquantified to show a clearer pattern in update times. 
% \hyperref[alg:generate_patch]{Algorithm~\ref{alg:generate_patch}} and~\hyperref[alg:apply_patch]{Algorithm~\ref{alg:apply_patch}} presents the pseudo code for creating and applying the patch.

% \begin{algorithm}
% \caption{Generate Model Patch}
% \label{alg:generate_patch}
% \begin{algorithmic}[1]
%     \State \textbf{Input:} Old model $\mathcal{M}$, New model $\mathcal{M}^*$
%     \State \textbf{Output:} Patch file $\mathcal{P}$
%     \vspace{5pt}

%     \State \textbf{Initialize} empty dictionary $\mathcal{P}$
    
%     \State \textbf{Iterate through parameters:}
%     \For{\textbf{each} parameter $p$ \textbf{in} $\mathcal{M}^*$}
%         \If{$p$ \textbf{not in} $\mathcal{M}$ \textbf{or} shape differs \textbf{or} values differ}
%             \State $\mathcal{P}[p] \gets \mathcal{M}^*[p]$
%         \EndIf
%     \EndFor
    
%     \State \textbf{Save} $\mathcal{P}$ to patch file

%     \State \textbf{Return} patch file $\mathcal{P}$
% \end{algorithmic}
% \end{algorithm}

\begin{algorithm}[t]
\caption{Apply Model Patch}
\label{alg:apply_patch}
\begin{algorithmic}[1]
    \State \textbf{Input:} Base model $\mathcal{M}$, Patch file $\mathcal{P}$
    \State \textbf{Output:} Updated model $\mathcal{M}^*$
    \vspace{5pt}

    % \State \textbf{Apply patch:}
    \For{\textbf{each} parameter $p$ \textbf{in} $\mathcal{P}$}
        \If{$p \in \mathcal{M}$ \textbf{and} $\text{shape}(\mathcal{M}[p]) = \text{shape}(\mathcal{P}[p])$}
            \State $\mathcal{M}[p] \gets \mathcal{M}[p] + \mathcal{P}[p]$ \Comment{Update existing parameters}
        \Else
            \State $\mathcal{M}[p] \gets \mathcal{P}[p]$ \Comment{Replace new or reshaped parameters}
        \EndIf
    \EndFor

    \State \textbf{Return} updated model $\mathcal{M}^*$
\end{algorithmic}
\vspace*{-0.2em}
\end{algorithm}

\section{Empirical Evaluation Methodology}\label{sec:experiment}

This section presents the methodology of the empirical evaluation of our proposed software update framework for heterogeneous UAV swarms. We investigate the following research questions~(RQs):

\begin{enumerate}[\bfseries RQ1. ]
\item \textbf{\rqone} 
The efficiency of update synchronization plays a crucial role in determining the optimal strategy for UAV swarms~\cite{Varadharajan2018}. A slow update process can lead to prolonged idling of the drones, hindering their ability to accomplish their task. Ensuring an efficient update process allows the UAV swarm to continue their ongoing tasks, which directly impacts the swarm's overall effectiveness and operational strength.
% , making it a critical consideration in update (synchronization) strategy selection.

% The efficiency of updates is an important factor in deciding which strategy is the optimal strategy to use. Ensuring an efficient update process directly impacts the swarm's ability to continue with their tasks. \cite{Varadharajan2018}.

\item \textbf{\rqtwo}
Overhead is another important factor in over-the-air software updates~\cite{8999425}, directly influencing system performance and efficiency. Overhead consists of communication signals exchanged and the resources consumed for packet transmission. Having excessive overhead can increase computational resource usage, consume more energy, and impact the speed of propagation. 

% Overhead is an important factor~\cite{8999425} in over-the-air software updates. Minimizing the overhead is important for the computational resource and the energy consumption.
% While UAV swarms can be considered inherently scalable due to their peer-to-peer (P2P) architecture \cite{Varadharajan2020}, it is important to measure scalability as some strategies may scale more effectively than others as the swarm size increases \cite{Mohammadi2019}.
\item \textbf{\rqthree}
Transferring a DL model across a UAV swarm to update to a new model version presents challenges, impacting both the update efficiency and transmission overhead consumption. High update size can cause high transmission costs, which can lead to prolonged update times, excessive overhead, and potential storage or communication bottlenecks, damaging the UAV swarm's performance.%  We investigate the trade-off of efficiency, overhead and accuracy when using frozen layers. %We use SwarmLLT with frozen layers to address this challenge by reducing the patch size while maintaining the functionality of the DL model.

% Transferring a large DL model will impact both the efficiency and the overhead consumption of the whole swarm. We introduce SwarmLLT to reduce the patch size and investigate the trade-off between efficiency, overhead, and accuracy.
% by applying deep learning strategies such as transfer learning\cite{torrey2010transfer}. Specifically, we freeze certain layers as described in \cite{9134370}.
% As UAVs are often used in mission-critical, business-critical, or safety-critical situations \cite{Xing2022}, ensuring that update strategies can tolerate failures and continue operating is important.

\end{enumerate}

\begin{algorithm}[t]
\caption{Gossip}
\label{alg:Gossip}
\begin{algorithmic}[1]
    \State \textbf{Receive} signal from Updater
    \State $PatchSize \gets$ from signal, propagate signal to neighbours
    \State \textbf{Receive} update packets
    \If{\#packets $<$ $PatchSize$} \Comment{Packet loss detected}
        \State \textbf{Request} retransmission
        \State \textbf{Receive} update packets
    \Else
        \State rebroadcasting $\gets$ \textbf{true}
    \EndIf
    \If{rebroadcasting}
        \For{each packet in update}
            \State \textbf{Send} packet
            \If{request received}
                \State \textbf{Retransmit} packets
            \Else
                \State \textbf{Signal} convergence
            \EndIf
        \EndFor
    \EndIf
    \If{signal\_converged received}
        \State Pass it on
    \EndIf
    \If{all nodes signal\_converged}
        \State Converged
    \EndIf
\end{algorithmic}
\vspace*{-0.2em}
\end{algorithm}

\begin{figure*}[t]
     \centering
     \begin{subfigure}[b]{0.33\textwidth}
         \centering
         \includegraphics[width=\linewidth]{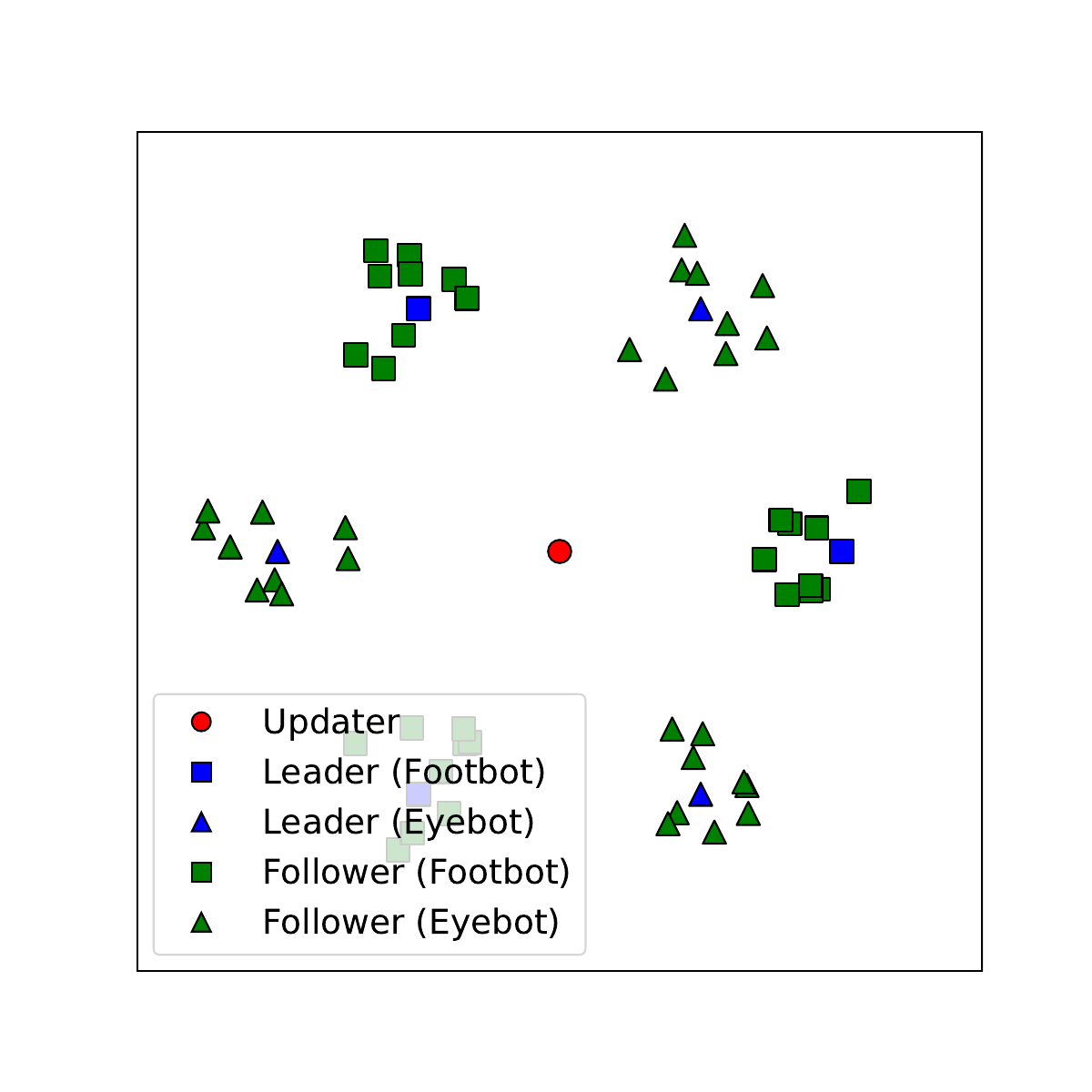} 
         \caption{SwarmSync}
         \label{fig:SwarmSync_structure}
     \end{subfigure}
     \hfill
     \begin{subfigure}[b]{0.33\textwidth}
         \centering
         \includegraphics[width=\linewidth]{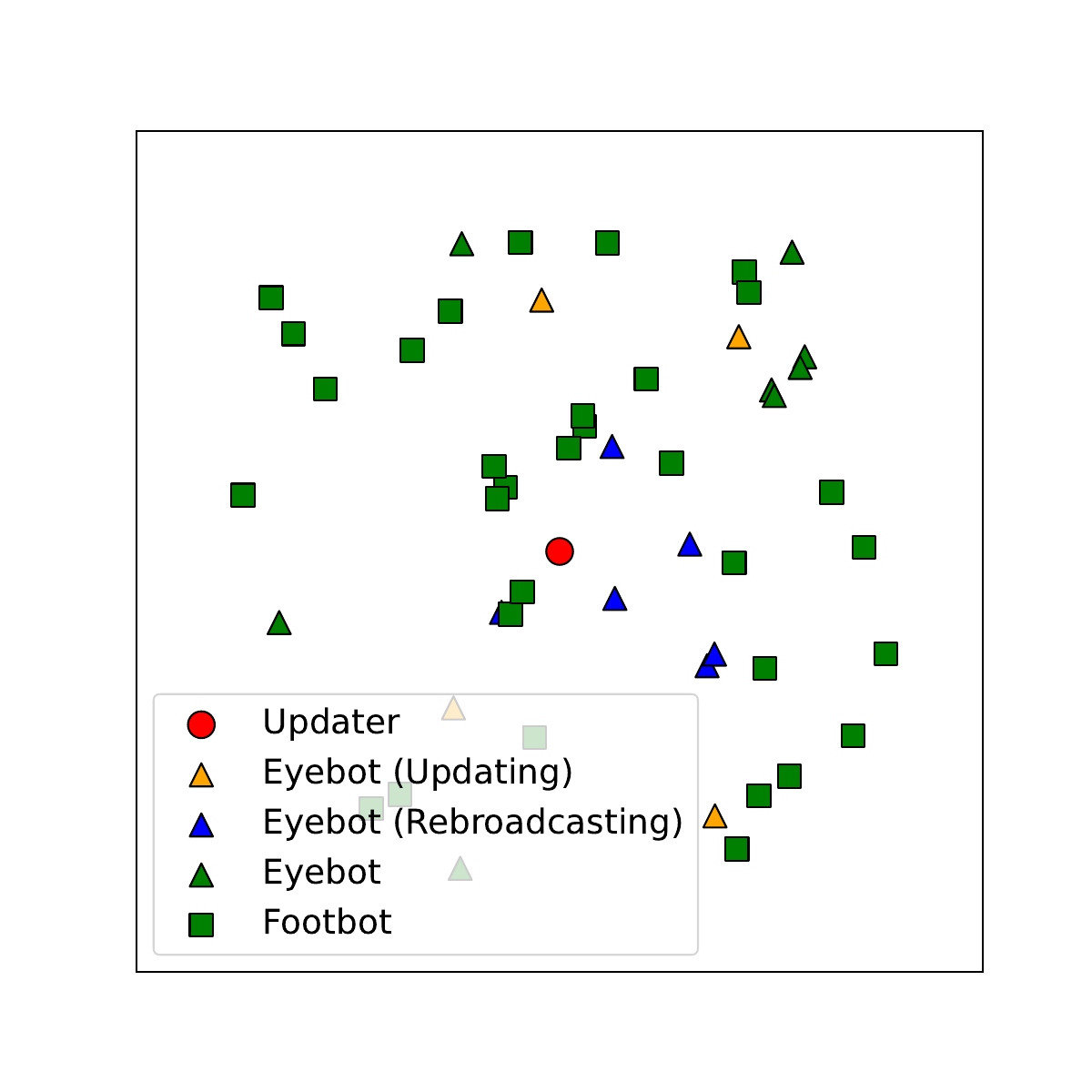} 
         \caption{Gossip}
         \label{fig:Gossip_structure}
     \end{subfigure}
     \hfill
     \begin{subfigure}[b]{0.33\textwidth}
         \centering
         \includegraphics[width=\linewidth]{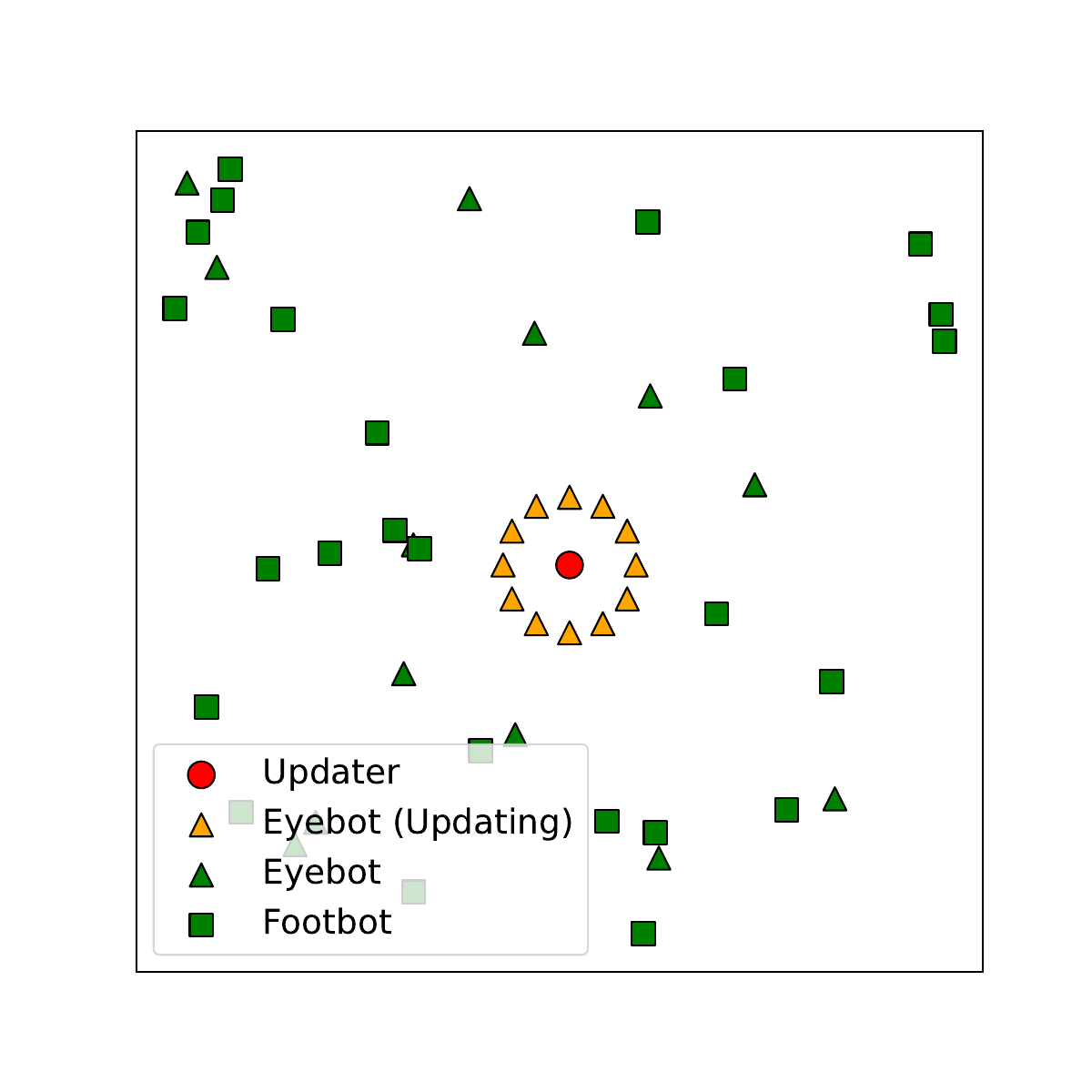} 
         \caption{SOUL}
         \label{fig:SOUL_structure}
     \end{subfigure}
\caption{Swarm UAV formations and roles for SwarmSync and the two baseline software update synchronization strategies.}
\label{fig:update_formation}
\vspace*{-0.5em}
\end{figure*}

\subsection{Baseline Strategies}

To empirically evaluate SwarmSync and SwarmModelPatch, we select state-of-the-art baseline techniques. For SwarmSync, we use a gossip-based rebroadcast baseline strategy and an auction-based baseline strategy (both illustrated in \autoref{fig:update_formation}), while for SwarmModelPatch we select a full-size update patch strategy.

\subsubsection{Gossip-Based Rebroadcast Strategy}

The gossip-based rebroadcasting method follows the principles of epidemic dissemination, modelled by the classic Susceptible-Infected-Recovered (SIR) epidemic model \cite{daley1964epidemics}. This model enables information to spread across the swarm with minimal time consumption, even in unreliable UAV communication networks where packet loss and intermittent connectivity are common. 

\citet{Varadharajan2018} uses such a gossip-based rebroadcasting approach to update a homogeneous swarm. In this strategy, if a receiver fails to compile the update due to incomplete packets, it requests retransmission from the sender, which will resend all packets. Once a drone successfully receives all packets, it transitions into an Updater, actively participating in further dissemination. The new Updater broadcasts the received packets to its neighboring drones, facilitating a decentralized propagation mechanism. 
However, as the Updaters are unaware of drones around them, they cannot determine when an update is finished. Gossip-based strategies can also generate high transmission overhead due to redundant rebroadcasting, making them inefficient for bandwidth constrained environments such as UAV networks. Additionally, these protocols do not differentiate between heterogeneous UAV types, leading to unnecessary information being propagated to drones that do not require them. 

As Gossip has already been implemented to update homogeneous swarms, we adapt it for heterogeneous swarms by introducing a simple modification. Specifically, drones that are not interested in the update will pass on the update message but will not receive or rebroadcast the update~(as shown in~\hyperref[alg:Gossip]{Algorithm~\ref{alg:Gossip}}). A timeout system similar is introduced: if none of the broadcasting drones receive an update request in 20 control steps of the simulator (see Section~\ref{sec:experimental-design}), the swarm is declared as converged.

% The algorithm for drones in Gossip is presented in~\hyperref[alg:Gossip]{Algorithm~\ref{alg:Gossip}}. An example structure for Gossip is shown in \autoref{fig:Gossip_structure}.

\begin{algorithm}[t]
\caption{SOUL's Updater distributing update}
\label{alg:SOUL_Updater}
\begin{algorithmic}[1]
    \State \textbf{Signal} update is available
    \State \textbf{Signal} update first group% \bram{sub-swarm?}
    \State \textbf{Wait} for signal UAVs are at location
    \State \textbf{Send} update
    \While{receiving requests}  \Comment{From UAVs}
        \State \textbf{Send} requested packets  \Comment{To UAVs}
    \EndWhile
    \State \textbf{Proceed} to next group% \bram{sub-swarm?}
    \If{all groups are done}
        \State \textbf{Converged}
    \EndIf
\end{algorithmic}
\vspace*{-0.2em}
\end{algorithm}

\subsubsection{Auction-Based Strategy (SOUL)}
% \begin{figure}[t]
%     \centering
%     \includegraphics[width=\linewidth]{images/Gossip_structure.pdf} 
%     \caption{An example structure formation for Gossip.}
%     \label{fig:Gossip_structure}
% \end{figure}

Auction-based strategies optimize data distribution and overhead by allowing UAVs to bid for content and only transmit necessary data. In this approach, the bidders need to bid for content that the auctioneer will send out after considering all the bids. While auction-based strategies minimize redundant transmissions, they are similar to a client-server model. As a result, when the number of bidders exceeds the maximum number of connections an auctioneer can handle, update times increase more linearly.
% While auction-based strategies have their advantages, being more similar to a client-server approach means that once the number of auctionees surpasses the max number of connections for the auctioneer, the time taken will grow more linearly.

SOUL~\cite{varadharajan2020soul} proposes an auction-based method for data sharing in a swarm. In SOUL, there is an auctioneer that handles the data blob, and it auctions the blob out for the swarm; any UAV in the swarm with available data storage and computation will bid on this blob. Based on the bidding, the auctioneer will decide how to split and distribute the blob to UAVs inside the swarm. Auction-based approaches allow updates to be distributed based on UAVs' availability, reducing redundant transmissions.

\begin{algorithm}[t]
\caption{SOUL's UAV receiving update}
\label{alg:SOUL drone}
\begin{algorithmic}[1]
    \State \textbf{Receive} signal from Updater
        \State $PatchSize \gets$ from signal, propagate signal to neighbours
    \State \textbf{Wait} for group's turn% \bram{sub-swarm?}
    \If{updating this group}
        \State \textbf{Move} to update location
        \State \textbf{Signal} at\_location
    \EndIf
    \State \textbf{Receive} packets \Comment{From Updater}
    \While{\#packets $<$ $PatchSize$}
        \State \textbf{Find} missing packets in self storage
        \State \textbf{Request} missing packets
        \State \textbf{Receive} packets
    \EndWhile
    \State \textbf{Signal} complete
    \State \textbf{Move} away from location
\end{algorithmic}
\vspace*{-0.2em}
\end{algorithm}

To adapt SOUL for updating a heterogeneous swarm, we set the Updater as the auctioneer and all UAVs as bidders. As shown in \hyperref[alg:SOUL_Updater]{Algorithm~\ref{alg:SOUL_Updater}} and \hyperref[alg:SOUL drone]{Algorithm~\ref{alg:SOUL drone}}, the Updater broadcasts the entire update, then UAVs that lack certain packets request retransmission via bidding. The Updater selectively rebroadcasts missing packets based on the received bids. UAVs whose positions are outside the Updater's range move toward the Updater to receive the update. If more than $N$ UAVs need the update, only $N$ will be allowed to go to the Updater and update at a given time. In this case, the swarm will form egalitarian groups and be updated group-by-group by the Updater. Similar to SwarmSync, SOUL drones will also propagate the update signal out to their neighbours. Since the Updater is unaware of the status of UAVs, if no request signal is received within a certain number of simulator control steps~(e.g., 20), the Updater assumes that the group has finished receiving the update. 
% An example SOUL formation is shown in \autoref{fig:SOUL_structure}

% This method can be adapted to software updates in a swarm. Specifically, the Updater in the swarm will broadcast the update to nearby drones, any drone that has missed a page will bid to the Updater. The Updater will re-transmit the pages based on the bidding received. The drones whose position is not in range of the Updater will path towards the Updater for the update. The Updater algorithm is presented in ~\hyperref[alg:SOUL_Updater]{Algorithm~\ref{alg:SOUL_Updater}}.

% To adapt this strategy to updates in UAV swarms, we made the following changes: The Updater will not split the data into blob but instead transfer the whole packet. The drones will not have to bid to the Updater with its storage and specifications, but rather bid for missing pages. The algorithm for drones in SOUL is presented in \hyperref[alg:SOUL drone]{Algorithm~\ref{alg:SOUL drone}}.

% As the Updater is not aware of the status of the drones, if no request signal is received in 20 control steps, the Updater will consider this group is finished.

% \begin{figure}[t]
%     \centering
%     \includegraphics[width=\linewidth]{images/SOUL_Structure.pdf} 
%     \caption{An example structure formation for SOUL.}
%     \label{fig:SOUL_structure}
% \end{figure}

\subsubsection{Baseline DL Model Patching Method}

No prior research has studied DL model patching for swarms. Since SwarmModelPatch offers adjustable model patch size by adjusting the number of frozen layers, a SwarmModelPatch variant that would not freeze any model layer during fine-tuning or re-training would be equivalent to fine-tuning or re-training the entire model, followed by replacing the previously installed model as a whole. Hence, we consider the SwarmModelPatch variant with zero frozen layers as our baseline, % as it yields a patch identical in size to the model itself, 
which we refer to as ``full-size update''.

% with more frozen layers offers smaller size and less frozen layers requires bigger patch size. We use SwarmLLT with zero frozen layers as our baseline to analyze the update strategies for the swarm, we refer to this as the full-size update. We use the full-size update as the baseline for it is the worse case scenario in terms of patch size.

% Also, for the results section, we have to specify that:
% In RQ1 and 2, we use the baseline DL model update with the proposed update strategies
% In RQ3, compare SwarmLLT with the baseline DL model update, also compare the results when integrating both

\subsection{Experimental Design}
\label{sec:experimental-design}

\subsubsection{Simulation Environment}

Our empirical evaluation uses ARGoS~\cite{Pinciroli2012} as the simulator to model swarm behaviour. ARGoS is a widely-used C++ simulator for robotics designed specifically for swarm simulation. %It supports multi-threaded execution for simulating large-scale swarm behaviours while maintaining computational efficiency.
While Gazebo~\cite{Koenig-2004-394} and Webots~\cite{Webots04} are also popular robotics simulators, ARGoS emphasizes multi-threaded execution, making it particularly suited for scenarios involving swarms with hundreds of UAVs. Furthermore, ARGoS was also used to evaluate SOUL and Gossip. ARGoS uses ``control steps'' as a measure of time and efficiency. In the default ARGoS configuration, each control step is 100ms, which we will use in our study to measure the efficiency of the synchronization and patch strategies. Each simulation is ran 10 times and taken the average to ensure the statical significance.% in control steps~(ticks).

\subsubsection{Update Configuration}

Similar to earlier research on heterogeneous swarms in ARGoS~\cite{ducatelle2010cooperative,ducatelle2011self}, we simulate a swarm consisting of three types of UAVs, each expecting different types of software updates: the Updater, Footbots, and Eyebots. Specifically, the Updater manages the update and relays it into the swarm, the Eyebots require a DL model update due to sudden weather changes~(from sunny to snowy), while the Footbot does not require any update. These updates symbolize a heterogeneous swarm where each type of UAV has its own update requirements. % \bram{any citation to show it's representative?}
As the ARGoS built-in ``eye-bot''\footnote{\url{https://pinciroli.net/api/dir_8bd2882cf9d14055c3b0116edcf0e192.php}} inherently cannot move, it does not represent a field-work use case, we use the ``foot-bot'' implementation to simulate a heterogeneous swarm composition, as ``foot-bot'' and ``eye-bot'' both share the same communication characteristics. We experiment with swarm sizes $s\in\{20, 100, 200, 500\}$ (excluding the Updater), with an equal ratio of $1:1$ between Footbots and Eyebots.

\subsubsection{SqueezeNet Model}

We choose SqueezeNet~\cite{iandola2016squeezenet} as the DL model for the Eyebots due to its lightweight architecture that modern drone hardware can handle~\cite{ms2022optimal}. For our evaluation, we select a 5-class weather classification task~\cite{alfaifi_5-class_weather_status_image_classification} to train the model. The dataset contains five classes: Sunny, Cloudy, Foggy, Rainy, and Snowy. To simulate environmental concept drift, we exclude the Snowy class from the original training data and consider Snowy as the new class in our simulation. After training SqueezeNet on the remaining 4 classes, we apply SwarmModelPatch to generate a model patch for a new model version that includes Snowy, with the replay buffer keeping 40\% of the old training data and combined with Snowy to obtain the final training data. As the dataset did not provide a balanced number of data per class nor was there inherent train/test seperation, we performed data augmentation to upscale all other classes to the class with the most data~(Sunny with 6274 images), then we split the data into train/test/val (0.7:0.15:0.15). Each model was trained 3 times and the best model is kept to study. To clearly analyze the impact of SwarmModelPatch on update size patterns and times, we do not apply quantization and other compression techniques that could further reduce patch sizes.
% \bram{how much of the data is used here for training and how selected? any cross-validation or bootstrap performed to obtain multiple variants of the model?}

In SqueezeNet, convolutional layers are represented as ``fire modules'', where each of the eight fire modules consists of multiple layers. For this study, we treat each fire module as a convolutional ``layer'' that could potentially be frozen. SqueezeNet also uses a convolutional layer as the classifier instead of a feed-forward neural network~(FNN); this layer is unfrozen by default. 

% We use the excluded class as the main training data for the updated model, adding a portion of the old training data to avoid catastrophic forgetting.~\cite{olewicki2024costs}. Specifically, we exclude snowy data to signify the change of season to winter. The data is pre-processed to meet the standard SqueezeNet requirements. 

% Training data will be portioned similarly to Lifelong Learning \cite{olewicki2024costs,parisi2019continual}. Lifelong learning is a method to keep training the deep learning model with new information, keeping a part of the older training data to be added to the new training data to ensure over-fitting does not occur on new data. For our experiment, we chose RBSize=50\%, higher than the original paper due to our model being smaller and more prone to over-fitting and catastrophic forgetting.  

\subsubsection{Communication Range}

\begin{figure}[t]
\centering
\includegraphics[width=0.65\linewidth]{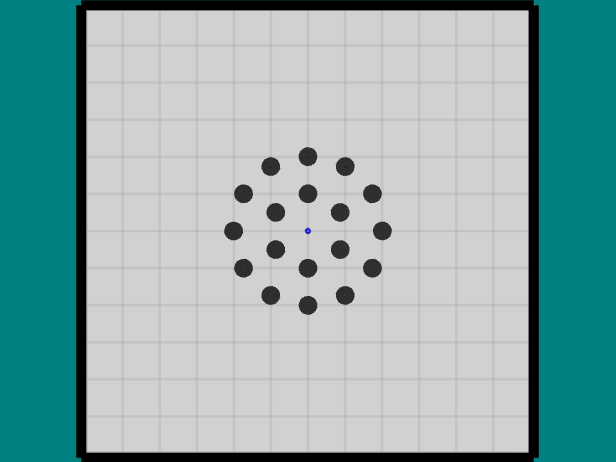} 
\caption{18 Drones surrounding the Updater.}
\label{fig:Example_formation}
\vspace*{-0.5em}
\end{figure}

As there are physical limitations for the physical formation of drones~(e.g., two drones cannot move into the same location), and ARGoS used a default communication length of $r=3m$, we standardize the maximum number of drones simultaneously updating around the Updater to $N=18$. This number is derived using a hexagonal distribution of drones and ARGoS' communication range, as shown in \autoref{fig:Example_formation}.% \bram{move that figure here instead of earlier in the paper?}. 

% Depending on the communication range $r$, the number of drones can be updated at a given time would change as well.

\subsubsection{Communication Failure}

We implement packet-loss scenarios as an all-or-nothing model using the ARGoS default configuration, meaning that if a packet is determined to be lost, the receiver will receive nothing. We use the same packet failure rates $f\in\{0,0.25,0.50,0.75\}$ as a previous study on homogeneous swarms~\cite{Varadharajan2018}.

% Based on a previous study on over-the-air-updates for homogeneous robotic swarms~\cite{Varadharajan2018}, we consider packet failure rates $f\in\{0,0.25,0.50,0.75\}$.

% chance, indicated by the failure rate $f$. For our experiment, we model packet failures as an all-or-nothing drop model aligning with ARGoS base configuration, we acknowledge that real-world UAV networks may experience losses differently. The failure rate is a percent measured $0\leq f \leq 1$, for our experiment, we only test for a subset of $f$. 
% Specifically, for RQ1 and RQ2, we use $f\in\{0,0.25,0.50,0.75\}$, for RQ3, we use $f=0.25$.

% The chosen values for $f\in\{0,0.25,0.50,0.75\}$ is derived from previous study on over-the-air-updates for robotic swarms~\cite{Varadharajan2018}. We chose to only keep one failure rate for RQ3 for simplicity, as we expect a similar pattern to RQ1 and RQ2 to occur if we add more failure rates. We chose $f=0.25$ for RQ3 as it is a common failure rate for sophisticated UAV systems \cite{petritoli2018reliability}.

\subsubsection{Transmission Speed}

While ARGoS' default data transmission speed is 100 bytes/sec, this is an artificial constraint that does not reflect real-world UAV communication speeds. We instead set a transmission speed of 1 Mbps based on the Bitcraze Crazyradio 2.0 module,\footnote{\url{https://www.bitcraze.io/products/crazyradio-2-0}} which is widely used in swarm robotics research. 
Since the default ARGoS configuration defines one control step as 100 ms, this means there are 10 control steps~(i.e., 10 packets) per second. To achieve a 1 Mbps transmission speed (equivalent to 125 kB per second), we set the packet size to 12.5 kB, ensuring that 10 packets per second result in the desired throughput.

\subsubsection{Swarm Behaviour}

In our simulation, the Updater will be deployed and remain stationary at $(0,0)$, while all Footbots and Eyebots are distributed randomly following a diffusion pattern for surveillance~\cite{ARGOS3Examples}. By default, the topology of the swarm is the cluster topology~\cite{pinciroli2016tuple} and the maximum speed for the drones is $V=1 m/s$, aligning with previous research~\cite{stolfi2024argos}.

\section{Results}\label{sec:results}

\subsection{RQ1: \rqone}\label{sec:rq1}

% \subsubsection{Setup}
% We evaluate the efficiency of the update strategies on swarm sizes $s\in\{20, 100, 200, 500\}$, with an equal ratio of $1:1$ between footbots and eyebots. We set the failure rates as $f\in\{0,0.25,0.50,0.75\}$. We measure the time taken from the start of the updates until the swarm converges.

\emph{Setup.}
To compare the efficiency of SwarmSync's with the two baseline update synchronization strategies, we use the baseline DL model update method~(``full-size update'') instead of SwarmModelPatch (which is evaluated in RQ3). As such, model updates correspond to 240 packets each. The efficiency of update synchronization is measured as the number of control steps from the start of the update synchronization until the swarm declares convergence.% \bram{this includes the full-size update done by the followers?}

%     \begin{figure}[h]
%     \centering
%     \includegraphics[width=0.3\textwidth]{images/E_0_RQ1.png} 
%     \caption{Gossip update process(1 drone).}
%     \label{fig:RQ1_0}
% \end{figure}
%     \begin{figure}[h]
%     \centering
%     \includegraphics[width=0.3\textwidth]{images/E_0.25_RQ1.png} 
%     \caption{Gossip update process(1 drone).}
%     \label{fig:RQ1_0.25}
% \end{figure}
%     \begin{figure}[h]
%     \centering
%     \includegraphics[width=0.3\textwidth]{images/E_0.5_RQ1.png} 
%     \caption{Gossip update process(1 drone).}
%     \label{fig:RQ1_0.5}
% \end{figure}
%     \begin{figure}[h]
%     \centering
%     \includegraphics[width=0.3\textwidth]{images/E_0.75_RQ1.png} 
%     \caption{Gossip update process(1 drone).}
%     \label{fig:RQ1_0.25}
% \end{figure}

% \begin{figure}[th]
%     \centering
%     \includegraphics[width=\linewidth]{images/RQ1.png} 
%     \caption{The average time taken for each strategy to converge the swarm.}
%     \label{fig:RQ1}
% \end{figure}

\begin{figure}[t]
    \centering
    \includegraphics[width=\linewidth]{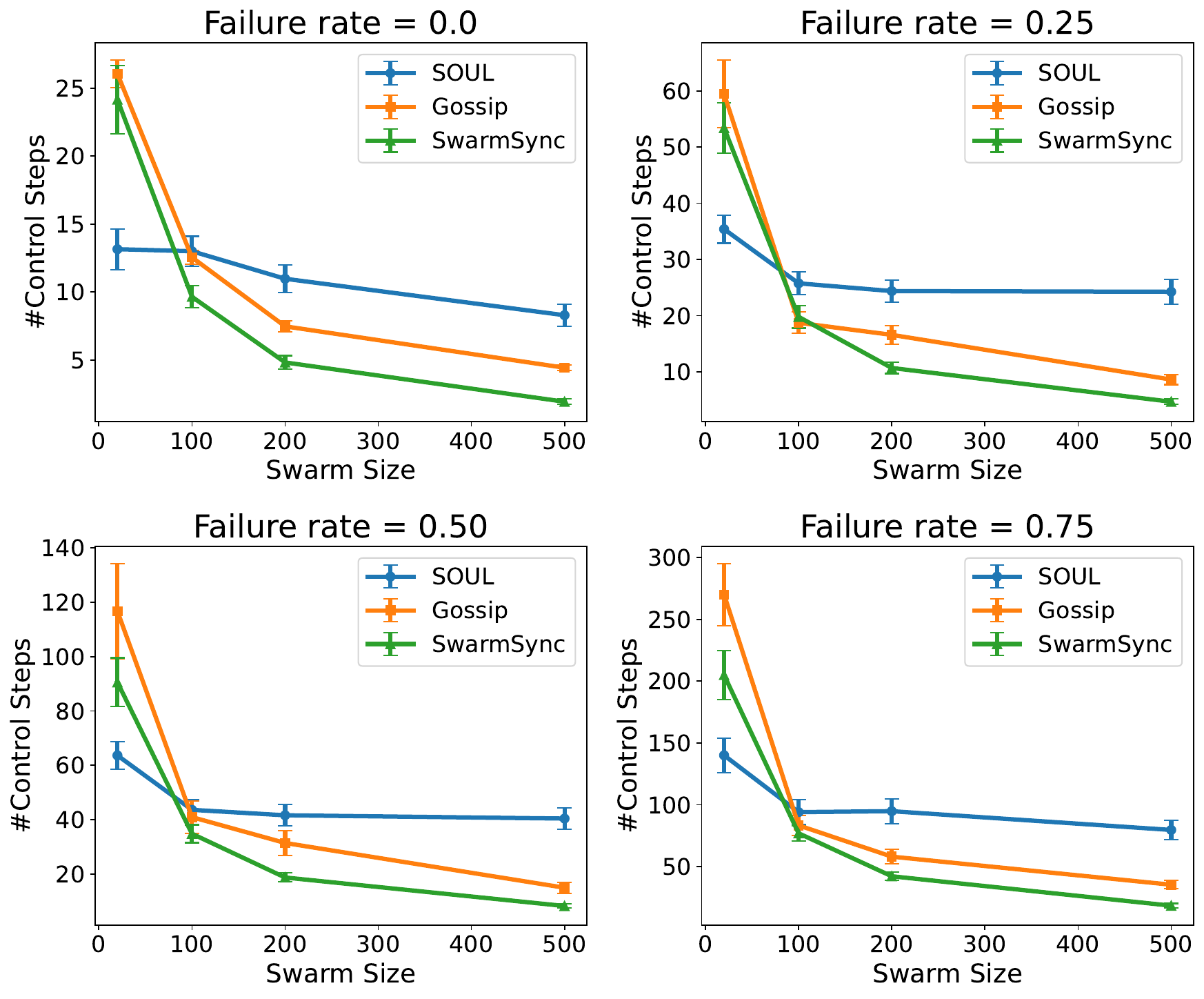} 
    \caption{The average time taken per drone for each synchronization strategy to converge the swarm.}
    \label{fig:RQ1_2}
\vspace*{-0.5em}
\end{figure}

\subsubsection{Findings}
\textbf{SwarmSync achieves the fastest convergence time across all strategies, outperforming SOUL by 78.3\% and Gossip by 47.7\% on average for the largest swarm size of 500.} \autoref{fig:RQ1_2} presents the average time taken across all drones to converge for varying swarm sizes. Compared to SOUL and Gossip, SwarmSync demonstrates efficient update propagation. At the largest swarm size 500, across all failure rates, SwarmSync completes the update on average 78.3\% faster compared to SOUL, and on average 47.7\% faster compared to Gossip. These results reinforce that SwarmSync is an efficient strategy for updating UAV swarms. However, SwarmSync shows minor inefficiencies for small swarms~(e.g., 20), which is likely due to a limitation in ARGoS where packets and their ACKs cannot be handled in the same control step, leading to wasted control steps that would not occur in real-life settings, this is discussed more in Section~\ref{sec:threats}.% \bram{which would not happen in real-life?}
% The ability to complete updates in 2 update cycles for swarm sizes $u \leq N^2$ makes SwarmSync a practical choice for large-scale UAV operations. For small swarm sizes~(e.g., $u=10$), SwarmSync updates slightly slower than SOUL. This is likely due to the limitation in ARGoS, where packets and their acknowledgements~(ACKs) are handled in the same control step, leading to wasted control steps.

\textbf{SwarmSync scales efficiently with increasing swarm sizes.} As the swarm size increases from 100 to 500 for failure rate 0.0, the average number of control steps decreased from 9.7 to 1.9 per drone for SwarmSync, showing a reduction by a factor of 5.1. This means that SwarmSync decreases the update synchronization time per drone approximately inversely proportional to the size increase; when the swarm size increased by a factor 5 times, the time per drone decreases by a similar factor of 5.1, implying more predictable and efficient scaling. SOUL exhibits higher convergence times across all failure rates, showing a time per drone trend almost parallel to the x-axis after it almost stabilizes after swarm size 100. While Gossip performs better than SOUL, it still shows a higher increase in convergence time as swarm sizes increase.

\subsection{RQ2: \rqtwo}\label{sec:rq2}

\subsubsection{Setup}
Using the same setup as RQ1, we evaluate the transmission overhead of the updates, which is accounted for when a drone sends out a packet~(12.5 kb), and when a drone sends out a communication signal (1 byte to 10 bytes), the average size of communication signal is 5 bytes. We measure the total overhead from the start of the update synchronization until the swarm converges
%, which is measured by the total amount of data sent during the updates across all drones (
in Megabytes. The update payloads remain the same as for RQ1.% \bram{what about communication signal size?}% Specifically, the overhead 

% The overhead will be measured in Mb, each packet is 12.5kb.

% \begin{figure}[th]
%     \centering
%     \includegraphics[width=\linewidth]{images/RQ2.png} 
%     \caption{The average overhead in Megabytes for each strategy to converge the swarm.}
%     \label{fig:RQ2}
% \end{figure}

\begin{figure}[t]
    \centering
    \includegraphics[width=\linewidth]{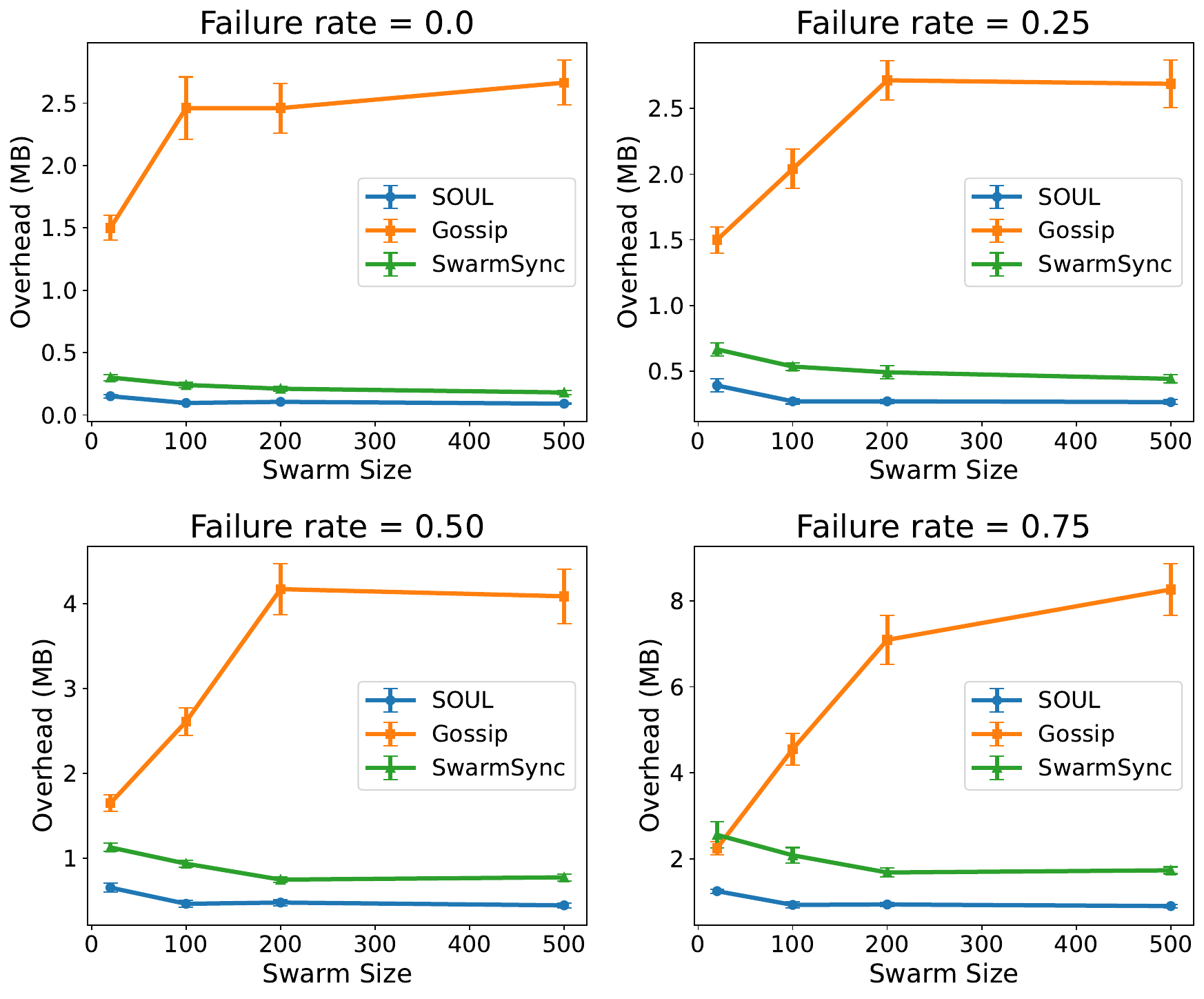} 
    \caption{The average overhead per drone for each synchronization strategy until convergence of the swarm.}
    \label{fig:RQ2_2}
\vspace*{-0.5em}
\end{figure}

\subsubsection{Findings}
\textbf{SwarmSync generates an average of 87.7\% more overhead than SOUL but only produces 19.7\% overhead compared to Gossip.} \autoref{fig:RQ2_2} shows, for both SwarmSync and SOUL, that transmission overhead decreases as swarm size increases. While SwarmSync produces an average of 87.7\% more overhead than SOUL across all failure rates and swarm sizes, the overhead for both SOUL and SwarmSync presents a decreasing trend as swarm size increases. In contrast, Gossip creates % more overhead as swarm size increases, by 
up to 8.3 MB of transmission overhead at a failure rate of 0.75, showing a 72.8\% increase in overhead compared to SwarmSync at a swarm size of 20.% The overhead of the Gossip strategy scales steeply as size increases.% This indicates that gossip has a high overhead cost as the system grows, making it less scalable for large swarms.
% \bram{it remains consistent and drops at the same time?} 

\textbf{Higher failure rates increase overhead across all three strategies, at swarm size 20, SOUL and SwarmSync show similar overhead growth (828.4\% and 850.5\%) as failure rate rises from 0.0 to 0.75.} As shown in \autoref{fig:RQ2_2}, the overhead for all strategies grows with increased failure rates across different swarm size settings. Notably, the relative overhead ratios among the strategies remain similar as the failure rate grows. Both SwarmSync and SOUL almost double in terms of overhead with each increase of the failure rate. Interestingly, while the overhead of both SwarmSync and SOUL grew by more than 100\% from failure rate 0 to 0.25, the overhead for Gossip remained relatively stable. 

However, both SOUL and SwarmSync maintain a relatively stable overhead increase~(around 15 times) when swarm size grows from 20 to 500 across all failure rates. The overhead for Gossip does not remain stable, with the same increase in swarm size, the increase in overhead goes from around 40 times for failure rate 0 and 0.25, to 62 times at failure rate 0.5, and 91 times at failure rate 0.75. This is likely due to the formation of Gossip, where lower sizes have multiple Updaters broadcasting to the same receivers, making the receiver get the update with minimal packet loss. With the size of the swarm increasing, the length between the Updater and the drones on the edge of the swarm increases, creating the need to get the update relayed through the entire swarm, growing in overhead.
% A similar pattern shows up in Gossip for a swarm size of 20 across all failure rates, with a higher increase only showing after 0.75. This is likely due to having a small number of drones, and almost all drones are inherently within range of the Updater, eliminating the need for UAVs to relay the update.
% From the graphs, we observe that while our method generates more overhead than SOUL, it is significantly more efficient compared to the Gossip(Broadcasting) strategy. 

%%% Local Variables:
%%% mode: LaTeX
%%% TeX-master: "../main"
%%% End:

\subsection{RQ3: \rqthree}\label{sec:rq3}

\begin{figure}[t]
    \centering
    \includegraphics[width=0.9\linewidth]{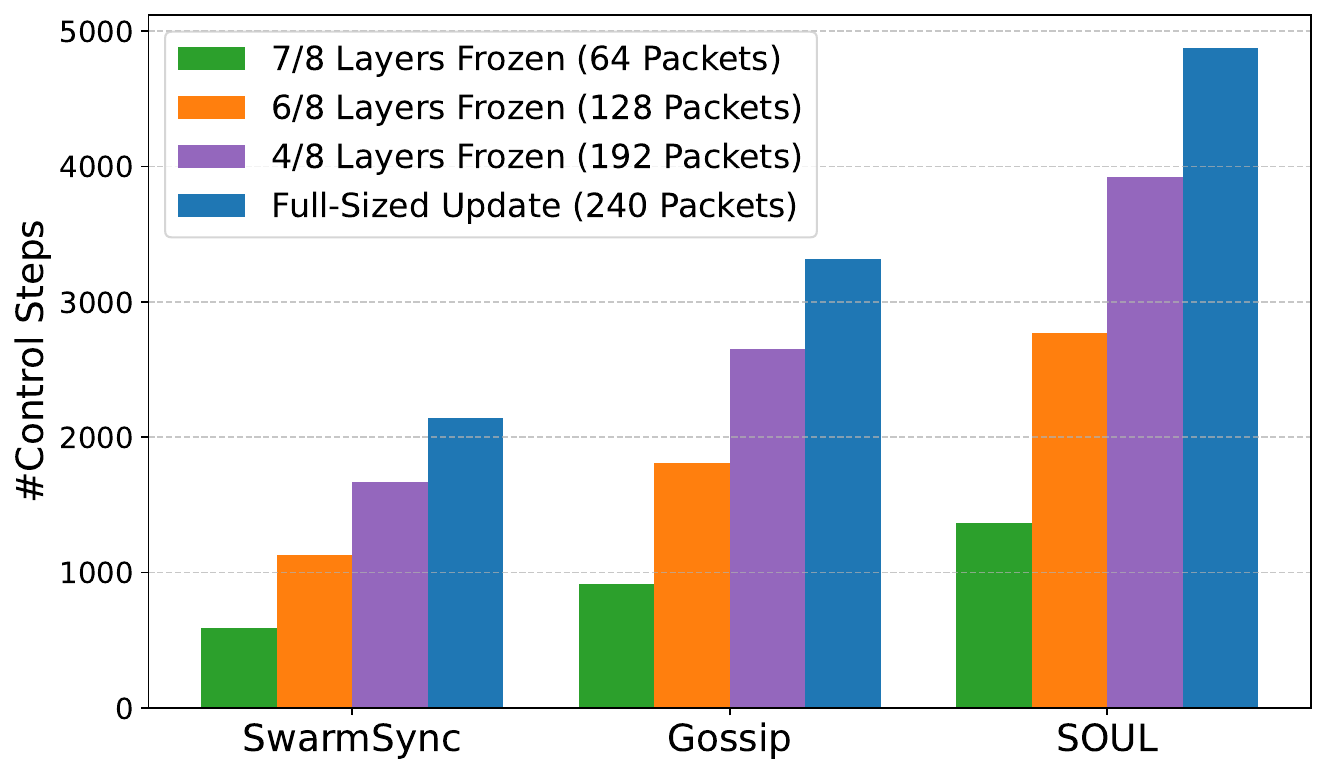} 
    \caption{The average number of control steps for swarm convergence across varying patch sizes.}
    %  (failure rate=0.25 and swarm size=200).
    \label{fig:E_RQ2}
\vspace*{-0.5em}
\end{figure}

\subsubsection{Setup}
% As updating the whole model is both high in overhead and time taken, we propose to only update certain layers to reduce the overhead and time taken. In situations where slight overall accuracy degradation is acceptable, this method will significantly improve the time to update the swarm.
To evaluate SwarmModelPatch's ability to reduce model patch size while maintaining model accuracy, % we investigate the impact of freezing different numbers of layers. Specifically, 
we progressively freeze the model layers up to the last layer~(``fire module'' in SqueezeNet), performing model updates for each number of frozen layers and comparing the trade-off between accuracy and update size with the ``full-size update'' baseline that was used in RQ1 and RQ2, each model is trained with 10 epochs. In SqueezeNet, layers closer to the classifier contain more weights, meaning that freezing those layers reduces the update size of the patch more than the other (more shallow) layers.

In particular, we select three configurations for our experiment: freezing 4, 6, and 7 fire modules. These configurations produce balanced reductions in patch size, with the configuration freezing 4 fire modules yielding a patch size approximately 300\% larger than the one obtained by freezing 7 modules, and freezing 6 fire modules resulting in a patch size approximately 200\% larger than the 7-module case. 
For example, the full-size update used in RQ1 and RQ2 requires 240 packets for one model version update. Freezing 4 fire modules reduces the model patch size to 192 packets, whereas freezing 6 and 7 fire modules further reduces the update sizes to 128 and 64 packets, respectively. We evaluate the updated models based on two metrics: overall classification accuracy (across all classes, including the new class data) and accuracy specifically on the newly introduced class data.% Overall accuracy is tested using a combination of old and new class data.

In addition, we evaluate the efficiency and overhead of different patch sizes in a swarm of 200 UAVs with a failure rate of $f=0.25$. We select swarm size = 200 to get 100 drones to be updated aligning with previous work~\cite{Varadharajan2018}. We select $f$ = 0.25 because it is a common failure rate for sophisticated UAV systems~\cite{petritoli2018reliability}. We measure the efficiency and overhead with the same measurements as RQ1 and RQ2. Similar to RQ1 and RQ2, we study all three strategies and the impact of patch sizes on them. 

\begin{figure}[t]
    \centering
    \includegraphics[width=0.9\linewidth]{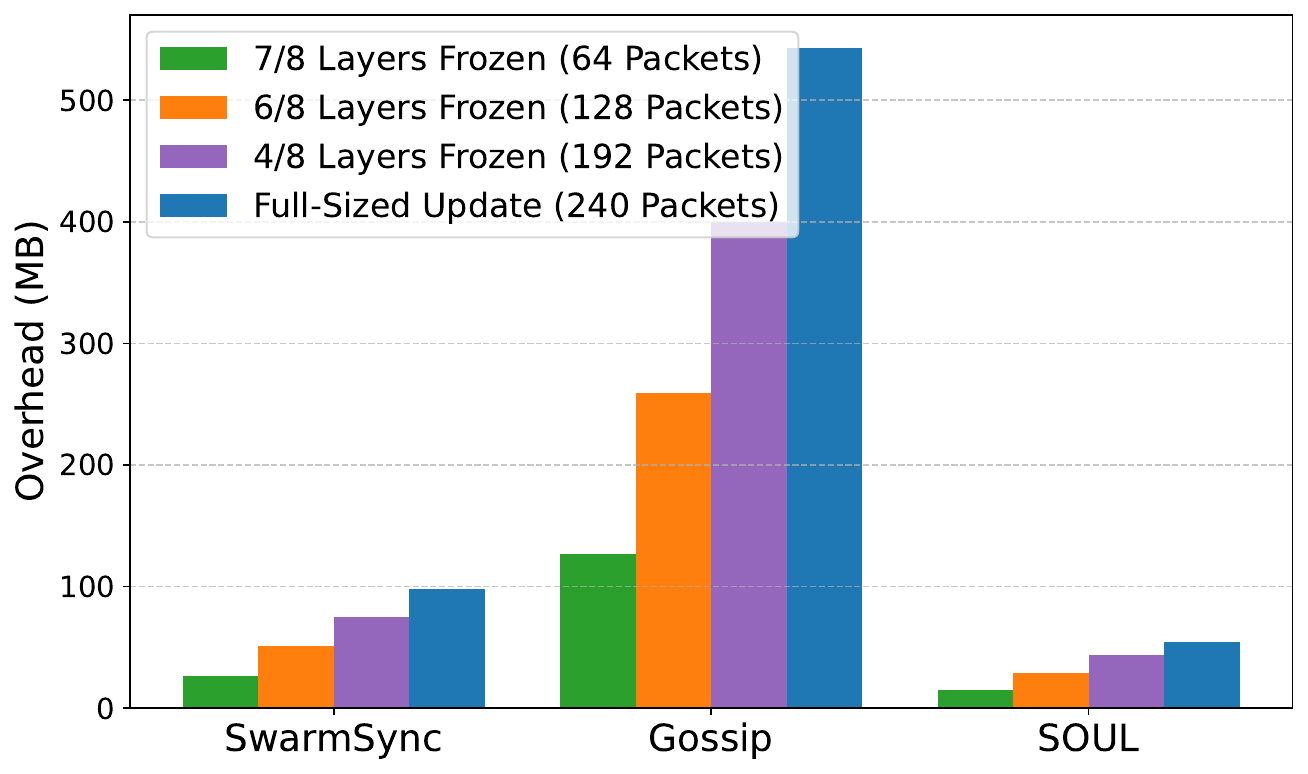} 
    \caption{The average overhead for swarm convergence across varying patch sizes.}
    %  (failure rate=0.25 and swarm size=200)
    \label{fig:O_RQ2}
\vspace*{-0.5em}
\end{figure}

\begin{table}[t]
\centering
\caption{Results of SwarmModelPatch for different numbers of frozen layers.}
\label{tab:model_comparison}
\begin{tabular}{lrrrr}
% \toprule
\hline
Model & \begin{tabular}[c]{@{}r@{}}Acc \\ (\%)\end{tabular} & \begin{tabular}[c]{@{}r@{}}New Class\\ Acc (\%)\end{tabular} & \begin{tabular}[c]{@{}r@{}}Size\\ (MB)\end{tabular} & \begin{tabular}[c]{@{}r@{}}Patch Size\\ (MB)\end{tabular} \\ \hline
0/8 Frozen (Baseline) & \textbf{72.7} & \textbf{72.3} & 2.9 & 2.9 \\
4/8 Frozen & 71.1 & 68.1 & 2.9 & 2.4 \\
6/8 Frozen & 68.6 & 67.4 & 2.9 & 1.5 \\
7/8 Frozen & 67.6 & 62.8 & 2.9 & \textbf{0.8} \\ \hline
% \bottomrule
\end{tabular}
\vspace*{-0.5em}
\end{table}

\subsubsection{Findings}
\textbf{SwarmModelPatch decreases patch size by up to 73.3\%, at the expense of a drop of 5.1\% in overall accuracy (from 72.7\%down to 67.6\%).}
\autoref{tab:model_comparison} presents the trade-off in accuracy when applying different numbers of frozen layers to SwarmModelPatch. As more layers get frozen, the overall accuracy lowers, from 72.7\% down to 67.6\% at 7/8 layers frozen, while the size of the patch decreases gradually from 2.9 MB down to 0.8 MB at 7/8 layers frozen. The new class~(Snowy)accuracy decreased by 9.5\% from 72.3\% down to 62.8\% at 7/8 frozen layers, showing a heavier impact than the overall accuracy. 
% The overall model size\bram{what does this mean:} stays constant at 2.9 MB, showing SwarmModelPatch does not affect the number of parameters in the models but only the patch size.

\textbf{SwarmModelPatch with frozen layers improves up to an average 72.3\% convergence speed at 7/8 layers frozen compared to full-size updates.}
\autoref{fig:E_RQ2} compares the number of control steps required for different update sizes under a failure rate of 0.25 with a swarm size of 200~(100 Eyebots and 100 Footbots). The results indicate that leveraging SwarmModelPatch with frozen layers improves convergence speed by an average of up to 72.3\%, with an error margin of 1\%, compared to updating the whole model.% The convergence speed for the three strategies in SwarmLLT updates resemble the same pattern as full-size updates. 

\textbf{SwarmModelPatch with frozen layers has up to an average 74.3\% lower transmission overhead at 7/8 layers frozen across all strategies.}
\autoref{fig:O_RQ2} presents the average transmission overhead (in Megabytes) evaluated under the same parameters as~\autoref{fig:E_RQ2}. Similar to~\autoref{fig:E_RQ2}, having a reduced size patch lowers the overhead transmitted during the update. The overhead for both SOUL and SwarmSync was reduced by up to 73.0\%(72.9\% and 73.0\%) for the 7/8 frozen layer update, while the overhead for Gossip was reduced by up to 76.7\%, which is likely due to UAVs in Gossip receiving packets from multiple senders at the same time.% keep the same pattern between the three strategies as the full-size update.

%%% Local Variables:
%%% mode: LaTeX
%%% TeX-master: "../main"
%%% End:

\section{Discussion}\label{sec:discussion}

\textbf{Selecting appropriate UAVs as leaders to improve overall update efficiency.}
In our simulation, UAV leaders are randomly assigned inside the sub-swarm since all UAVs are identical. In practice, having leaders equipped with more powerful physical capabilities could enhance the performance of the swarm~\citep{varadharajan2024hierarchies}. Practitioners should select leaders based on key performance indicators such as communication bandwidth, computational power, data storage capacity, and battery life. Prioritizing UAVs with enhanced resources as leaders could reduce synchronization time and increase update efficiency in real-world scenarios.

% \subsection{Implications for Researchers}

\textbf{Future work should investigate security threats when updating heterogeneous swarms.} 
Our study leverages a structured, hierarchical update mechanism designed for heterogeneous swarm scenarios. However, the security threats during updates in these heterogeneous settings remain unexplored. Although prior studies, such as \citet{al2022study}, have investigated security threats of software updates for homogeneous UAV swarms, no studies address security concerns specific to heterogeneous swarms. Software update for heterogeneous swarms could introduce unique attack vectors, where adversaries may target role-specific vulnerabilities. Hence, future research should focus on: (1)~understanding the vulnerabilities specific to heterogeneous swarms, and (2)~proposing approaches to enhance security~(e.g., role-based authentication).

% \textbf{Researchers should investigate security implications.} While our study does not explicitly address security concerns, future work could explore how security measures might interact with SwarmSync's synchronization strategy. For instance, implementing role-based validations for updates could mitigate the risk of malicious attackers stealing information or introducing software failures. We propose two directions: (1) investigate the trade-off between security implementations and update propagation performance, and (2) evaluating whether our observer pattern for the three strategies persists under adversarial conditions. By addressing these questions, researchers can extend our findings to practical settings while preserving robustness—a critical step toward real-world adoption.

\textbf{Future work should explore SwarmModelPatch in other resource-constraint scenarios.}
The proposed model patching algorithm, SwarmModelPatch, is specifically designed to tackle resource constraints inherent in UAV swarms scenarios, where limited bandwidth, computational capability, and energy constraints are major bottlenecks. SwarmModelPatch could be applied in other resource-constrained domains, for example, the Internet of Things (IoT) or edge computing platforms. Researchers should explore adapting and evaluating SwarmModelPatch in these scenarios.

\section{Threat to Validity}\label{sec:threats}
% In this section, we discuss the potential threats to validity of our empirical evaluation of SwarmSync and SwarmModelPatch on heterogeneous UAV swarms.
% about software update frameworks for
% Internal threats are issues with the simulator making potential deviations from an optimized simulation setup. External threats are potential problems when our strategies get adapted and implemented in a real-world setting.
% \textbf{Construct validity.}

\noindent\textbf{Internal Validity.}
% In ARGoS, a drone can send communication signals to all neighbors and receive communication signals from all neighbors, without impacting the speed of transmission/receive. This impacts Gossip the most as realistically UAVs cannot accomplish the same thing in a real-world setting.
ARGoS simulates time in discrete control steps, but its implementation of message transmission and reception introduces a limitation: the receiver can only read signals from the last control step. When a packet is sent at time $t$, the receiver can only receive the packet at the next control step $t+1$. Hence, even if the receiver would immediately send an ACK back to the sender, the sender would receive the ACK at time $t+2$. In an optimized control system, at failure rate $f=0$, when a packet is sent at time $t$, the ACK packet will also be sent back at time $t$, and the sender could move on to the next packet at time $t+1$, yet ARGoS currently does not allow this, making SwarmSync take longer in the simulation. This effect does not affect the other strategies, as other strategies are not reliant on a response before sending the next packet.

In our simulation, we used Footbots and Eyebots to model a heterogeneous UAV swarm, as this is a common setup in ARGoS for such studies. However, this choice may not fully reflect real-world heterogeneous swarm configurations. We maintained an equal ratio of Footbots and Eyebots, though practical applications may vary. Additionally, in ARGoS, we represented the drones' communication range as a circular area, whereas in real-world scenarios, this range is typically more spherical.

\noindent\textbf{External Validity.}
% As we have designed this experiment as a simulation, we acknowledge that there are more complexities when the findings are applied to the real world, as also discussed in the previous section.
In the simulation, it is assumed that all drones are connected to the swarm and that each drone has at least one relay path to receive information. While this is a common assumption for simulation~\cite{pinciroli2016tuple}, depending on the situation, this assumption might not hold true for real-world use cases. %In the case where some drones could be disconnected from the main swarm, \bram{the disconnected ones?} communication drones in the swarm would need to navigate based on their belief of where the separated sub-swarm could be, similar to a hide-and-seek problem in reinforcement learning. \bram{are there any approaches for doing that that we could cite? that would make it less of an issue}

In addition, the packet loss is modelled as an all-or-nothing drop. In real-world settings, this is likely not the case. Preventative measures such as hashing should be used to confirm the integrity of the packet. If hashing fails for a packet, then the packet is declared as corrupted, and should be requested again as if it was missing. We also acknowledge that in real-world implementations, a static 1Mb/s transfer speed is not realistic, adaptive bitrate selection may be needed based on interference and range limitations.

We evaluated SwarmModelPatch using SqueezeNet on the 5-class weather classification dataset~\cite{alfaifi_5-class_weather_status_image_classification}. Note that the trend of model performance observed with different numbers of frozen layers may differ when applying SwarmModelPath to other DL models or datasets.

% BATTERY

% \bram{need to mention our focus on 1 specific CNN model, our focus on one task and dataset, our focus on CNN models in general (instead of other DL types), our use of specific formations of drones, etc.}

%%% Local Variables:
%%% mode: LaTeX
%%% TeX-master: "../main"
%%% End:

\section{Conclusion}\label{sec:conclusion}

This study presents SwarmUpdate, the first framework for software updates in heterogeneous UAV swarms, comprising the SwarmSync update synchronization approach and SwarmModelPatch DL model patching approach.
We empirically evaluated SwarmSync by comparing it against two baseline update synchronization strategies~(SOUL and Gossip) and one baseline model update approach (full-size update) in heterogeneous swarm simulations. % By adapting and comparing these strategies, we compare their robustness, scalability, efficiency, and overhead trade-offs. 
Our results show that SwarmSync is the most efficient and scalable out of the three synchronization strategies, achieving up to 78.3\% faster convergence times than SOUL and 47.7\% faster convergence than Gossip, while maintaining a reasonable overhead of 19.7\% compared to Gossip. In addition, our findings confirm the viability of incremental model updates using SwarmModelPatch, showing how it can reduce update time and overhead by up to 72.3\% and 74.3\%, respectively, without substantially compromising model accuracy.  
% We showcase the trade-off of using a reduced-size patch to apply updates, we find that we were able to reduce the update time and overhead without significantly compromising model accuracy. 

This research lays a foundation for future studies on software update frameworks for heterogeneous swarm systems, enhancing their reliability and effectiveness. 
% Our findings contribute to understanding and improving software updates in heterogeneous, peer-to-peer UAV swarms.
As UAV technology continues to advance, ensuring robust, efficient, and scalable updates will be crucial for mission-critical and large-scale UAV deployments (e.g., humanitarian and surveillance missions).
% Future work should focus on the real-world implementation and validation of the findings presented in this paper. Exploring strategies that further optimize the update efficiency in dynamic, failure-prone environments.

%%% Local Variables:
%%% mode: LaTeX
%%% TeX-master: "../main"
%%% End:

\section{Acknowledgements}
We extend our gratitude to Vivek Shankar Varadharajan for his invaluable support in providing the initial guidelines and source codes for \cite{Varadharajan2018,varadharajan2020soul} that contributed to this work.

%%
%% The next two lines define the bibliography style to be used, and
%% the bibliography file.
\bibliographystyle{ACM-Reference-Format}
\bibliography{main}

\end{document}